\documentclass[aps,pra,twocolumn,footinbib,
showpacs,showkeys,longbibliography, nobalancelastpage]{revtex4-1}

\usepackage{graphicx}
\usepackage{indentfirst}
\usepackage{physics}
\usepackage{braket}
\usepackage{float}
\usepackage{amsmath}
\usepackage{amssymb}
\usepackage{verbatim}
\usepackage{wasysym}
\usepackage{CJK}
\usepackage{esint}
\usepackage{color}
\usepackage{xcolor}
\usepackage{subfigure}
\usepackage{amsfonts}
\usepackage{footmisc}
\usepackage{scrextend}
\usepackage{multirow}
\usepackage{mathtools}
\usepackage{xr-hyper}
\usepackage[hyperfootnotes=false]{hyperref}

\usepackage[english]{babel}
\usepackage{url}
\usepackage{bm}
\definecolor{darkblue}{rgb}{0,0,0.5}
\hypersetup{
colorlinks=true,
linkcolor=black,
filecolor=black,
citecolor=darkblue,
urlcolor=black,
}
\newcommand{\defeq}{\vcentcolon=}
\newcommand{\eqdef}{=\vcentcolon}

\DeclareMathOperator*{\argmin}{arg\,min}
\newcommand{\enn}{E_{\text{NN}}}

\newcommand\bs[1]{\boldsymbol{#1}}

\begin{document}

\title{Ultimate Limits of Thermal Pattern Recognition}

\author{Cillian Harney${}^{1}$}
\author{Leonardo Banchi${}^{2,3}$}
\author{Stefano Pirandola${}^{1}$}
\affiliation{${}^{1}$Department of Computer Science, University of York, York YO10 5GH, UK}
\affiliation{${}^{2}$Department of Physics and Astronomy, University of Florence,
via G. Sansone 1, I-50019 Sesto Fiorentino (FI), Italy}
\affiliation{${}^{3}$INFN Sezione di Firenze, via G. Sansone 1, I-50019, Sesto Fiorentino (FI), Italy}

\begin{abstract}
Quantum Channel Discrimination (QCD) presents a fundamental task in quantum information theory, with critical applications in quantum reading, illumination, data-readout and more. The extension to multiple quantum channel discrimination has seen a recent focus to characterise potential quantum advantage associated with quantum-enhanced discriminatory protocols. In this paper, we study thermal imaging as an environment localisation task, in which thermal images are modelled as ensembles of Gaussian phase insensitive channels with identical transmissivity, and pixels possess properties according to background (cold) or target (warm) thermal channels. Via the teleportation stretching of adaptive quantum protocols, we derive ultimate limits on the precision of pattern classification of abstract, binary thermal image spaces, and show that quantum-enhanced strategies may be used to provide significant quantum advantage over known optimal classical strategies. The environmental conditions and necessary resources for which advantage may be obtained are studied and discussed. We then numerically investigate the use of quantum-enhanced statistical classifiers, where quantum sensors are used in conjunction with machine learning image classification methods. Proving definitive advantage in the low-loss regime, this work motivates the use of quantum-enhanced sources for short-range thermal imaging and detection techniques for future quantum technologies.
\end{abstract}

\maketitle

The core aspiration of developing future quantum technologies is to exploit the intrinsic features of quantum mechanics in order to outperform any known optimal classical strategy for a given problem. Amidst rapidly accelerating progress in a number of fields of quantum computing \cite{Mike_Ike, preskillnisq, GoogleSyc}, communications \cite{AdvCrypt} and more, quantum sensing stands as the most mature and deployable of these fields \cite{QSensing}. This can be attributed, in large part, to a collection of theoretical advances in quantum decision theory/hypothesis testing \cite{HelstromQHT, BaeQSD, QSDBergou, QSDChefles, DiscrimQOps}, quantum metrology \cite{QMetr}, and a focus on Continuous Variable (CV) bosonic quantum information which is particularly apt for quantum sensing \cite{AdvPhS, GaussRev, SerafiniCV}.\par
An important setting of quantum hypothesis testing for a number of quantum technologies is Quantum Channel Discrimination (QCD), in which a user is tasked with classifying between a number of quantum channels using a (possibly) quantum-enhanced discrimination protocol. QCD finds crucial applications within quantum illumination \cite{Lloyd2008, Tan2008, Shapiro2009, Zhang2013, Zhang2015, Barzanjeh2015, Dallarno2012, Zhuang2017_1, Zhuang2017_2, LasHeras2017, DePalma2018, NairIllum, BarelySep}, quantum reading \cite{PirBD, NairNDS,Tej2013, Spedalieri2012, GSCryptoQR, Zhuang2017_3,Hirota2017}. Here, major quantum advantage has been theorised and in some cases experimentally verified \cite{Lopaeva2013, Barzanjeh2020, expqread, QEnhDataCl}, by exploiting non-classical properties of quantum input states and measurements.\par
Yet in general, this is a very difficult double optimisation problem, as one must determine the optimal input (or probe) state and optimal measurement used to minimise the error probability of misclassification. Furthermore, the most general protocol will make use of adaptive operations, which are extremely difficult to characterise and optimise. Nonetheless, critical performative insight into the fundamental lower bounds of discrimination error probabilities have been unveiled through the employment of channel simulation \cite{FLQCD,UltPrec2017,TCS}. By simulating channels through the use of quantum programmable processors, one can reduce channel discrimination to state discrimination and allow for the derivation of ultimate lower bounds. These methods are extremely powerful and have been used in a plethora of open problems in quantum communications, sensing and machine learning \cite{PLOB,TserkisEC,Laurenza2018,Laurenza2020,ConvBanchi}.\par
While the majority of QCD research has focussed on binary discrimination tasks, recent advances in the realm of multi-channel discrimination via the formulation of Channel Position Finding (CPF) \cite{EntEnhanced, ULMCD} has invited new exploration into the multi-ary domain. In this context, one may model images as consistent of pixels described by a target or a background quantum channel. This introduces a pattern recognition problem in which the goal is to perfectly classify a distribution of channels across an image. \par
Introduced in the context of barcode discrimination \cite{PatternRecog} in which pixels are modelled as pure lossy channels (relevant in the optical regime), these ideas can be extended into a thermal imaging setting. That is, thermal images may be described as a collection of pixels which are modelled by Gaussian phase insensitive channels subject to thermal noise. Formulated recently by Pereira et al.\ \cite{OptEnvLoc} in the context of single CPF, discrimination then becomes an environment localisation task in which one must discriminate between target/background channels of identical loss but different noise properties. Importantly, these collections of channels exhibit a property called joint teleportation-covariance, meaning one may employ multi-channel simulation techniques in order to derive ultimate error probability lower bounds. This presents a fascinating opportunity to unveil the optimal performance of quantum-enhanced thermal imaging, and provide insight into its future applications and limitations.\par
This paper proceeds as follows: In Section \ref{sec:TeleCov} we review the image formulation introduced in \cite{PatternRecog, OptEnvLoc} in the context of environment localisation and the application of teleportation stretching that allows for the determination of ultimate lower bounds. In Section \ref{sec:ExactClass} we begin with the fundamental binary discrimination problem (a single pixel) in the context of thermal imaging, laying the foundations for exact pattern classification. Section \ref{sec:SLearn} then makes use of statistical learning techniques in which a sensor is assisted by supervised learning, which allows us to explore larger image dimensions with lower resource demands.

\section{Teleportation Stretching \& Thermal Imaging\label{sec:TeleCov}}
\subsection{Thermal Channel Patterns}
Throughout this work we consider the following scenario: A thermal image $\bs{i}$ consists of $m$-pixels where each pixel is characterised by a Gaussian phase insensitive channel with identical transmissivities $\tau$, but different environmental noise $\nu$ corresponding to a background channel $\mathcal{E}_B$ or a target channel $\mathcal{E}_T$. For instance, target channels may correspond to warmer pixels that we may wish to identify against a background of colder pixels. 
An $m$-pixel image can therefore be modelled as a channel pattern $\mathcal{E}_{\bs{i}}^m  \defeq  \bigotimes_{k=1}^m \mathcal{E}_{i_k}$ where $i_k \in \{B,T\}$. Let
\begin{equation}
I_{[x]_{\bs{i}}}^m \defeq  \bigoplus_{k=1}^m  x_{{i_k}} I ,
\end{equation}
be a function of a channel property $x$ dependent on the position $k$ in a pattern $\bs{i}$, where $I$ is the $2 \times 2$ Identity matrix. Assuming an input $m$-mode Gaussian state fully characterised by its Covariance Matrix (CM) (with zero first moments), this multi-channel will perform the following transformation on its CM,
\begin{equation}
V_{\bs{i}} =( I_{[\sqrt{\tau}]_{\bs{i}}}^m) V  ( I_{[\sqrt{\tau}]_{\bs{i}}}^m)^T +  I_{[\nu]_{\bs{i}}}. \label{eq:CMtrans}
\end{equation}
Thermal-loss/amplifier channels are modelled by a beamsplitter of transmissivity $\tau$, mixing the input signal with an environmental mode described by a thermal state of mean photon number $\bar{n}$. We consider there to exist background and target channels with equivalent transmissivities but different thermal numbers $\bar{n}_j$ for $j\in \{B,T\}$. Thermal-loss (amplifier) channels are parameterised by $0 \leq \tau < 1$ ($\tau > 1$) and $\nu \geq \frac{|1-\tau|}{2}$ for shot-noise $= \frac{1}{2}$. The induced noise of the target and background channels are related to the local thermal numbers in each mode as 
\begin{gather}
\nu_{j} = \epsilon_j |1-\tau|,\text{ such that } \epsilon_j \defeq \bar{n}_j + \frac{1}{2}.
\end{gather}
Meanwhile, Gaussian additive-noise channels are parameterised by $\tau = 1$ and $\nu_j \geq 0$, i.e.\ they are ideal channels with zero-loss, but are still subject to non-zero noise.\par
An ensemble of possible patterns occurring with probability $\pi_{\bs{i}}$ can define an image space $\{\pi_{\bs{i}}, \mathcal{E}_{\bs{i}}^m\} =  \mathcal{U}$. Let us define some key binary image spaces: The set of all $2^m$ possible uniform channel patterns, with no global restrictions, is given by $\mathcal{U}_{m}$, studied as barcodes in \cite{PatternRecog}. If we restrict the space to the set of all $m$-channel patterns containing $k$ target channels, then we call this problem $k$-Channel Position Finding ($k$-CPF) \cite{EntEnhanced} since we are trying to locate the positions of these $k$ channels. This image space is denoted by  $\mathcal{U}_{\textsf{\tiny CPF}}^k$. More generally, we can define some fixed set of target channel numbers $\bs{k}$ characterising bounded $k$-CPF, $\mathcal{U}_{\textsf{\tiny CPF}}^{\bs{k}}$, such that there is uncertainty over the exact number of target channels present. Note that
\begin{equation}
\mathcal{U}_{\textsf{\tiny CPF}}^k \subseteq \mathcal{U}_{\textsf{\tiny CPF}}^{\bs{k}} \subseteq \mathcal{U}_m,
\end{equation}
defines a hierarchy of image spaces, and the larger the image space, the more difficult discrimination becomes. 

\begin{figure*}[t!]
\includegraphics[width=\linewidth]{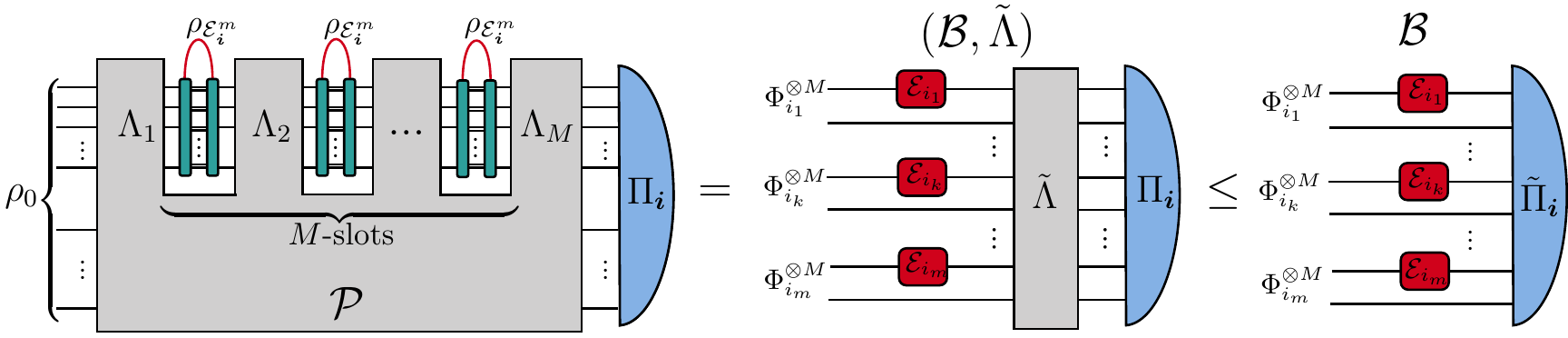}\\
\hspace{2.5cm} (a) \hspace{7.1cm} (b) \hspace{3.8cm} (c) 
\caption{Panels (a)-(c) depict the teleportation stretching of a general adaptive protocol, described by a quantum comb $\mathcal{P}$ that contains teleportation covariant channels.  Via teleportation stretching, the general adaptive protocol is stretched to (b) a block-assisted protocol with the generation of $M$-copy Choi states, followed by a pre-measurement adaptive operation, $(\mathcal{B},\tilde{\Lambda})$. The performance of (a) and (b) can then be bounded by the block-assisted protocol (c) in the limit of infinitely squeezed TMSV states in a probe/idler configuration, which achieves the ultimate performance limit. 
}
\label{fig:adaptiveprotocol}
\end{figure*}

\subsection{Multi-Channel Discrimination Protocols}
The most general adaptive discrimination protocol is best described using a quantum comb \cite{Laurenza2018, ChiribellaComb} (see Fig.~\ref{fig:adaptiveprotocol}(a)). This is a quantum circuit board that possesses $M$ ``slots"  which may be filled with instances of $m$-length channel patterns $\mathcal{E}_{\bs{i}}^m$. The comb itself is a register of an arbitrary number of quantum systems, initially prepared in some state $\rho_{0}$. Each channel pattern instance offers the opportunity to interrogate the multi-channel using some quantum systems within the register. Before and after each pattern instance, the discriminator may perform arbitrary joint quantum operations (QOs), which can be assumed to be trace-preserving \footnote{Since we have an unlimited number of systems in the register, and thanks to the principle of deferred measurement}. This allows for the exploitation of unlimited entanglement, shared between input and output and feedback that can be used to adaptively optimise subsequent pattern interactions. After $M$ adaptive probings, the comb is in its final state $\rho_{\bs{i}}^{M}$, and is subject to an optimal, joint POVM $\{\Pi_{\bs{i}^{\prime} }\}_{\bs{i}^{\prime} \in \mathcal{U}}$ which is classically post-processed to infer the channel pattern with some probability of error. Let us denote general adaptive protocols by $\mathcal{P}$, with classification error probability 
\begin{equation}
p_{\text{err}}(\mathcal{P}) \defeq  \sum_{\bs{i} \neq \bs{i}^{\prime}} \pi_{\bs{i}} \text{Tr}\left[ {\Pi}_{\bs{i}^{\prime}} \rho_{\bs{i}}^{M}\right] . \label{eq:p_err_gen}
\end{equation}
where $\sum_{\bs{i}\neq \bs{i}^{\prime}}$ is a sum over all unequal channel patterns in the image space. \par
We may instead specify our strategy and consider block/block-assisted protocols, $\mathcal{B} \subset \mathcal{P}$. Here, a channel pattern $\mathcal{E}_{\bs{i}}^m$ is probed identically and independently by $M$-copies of an input state $\rho^{\otimes M} \rightarrow \rho_{\bs{i}}^{\otimes M} \defeq \mathcal{E}_{\bs{i}}^m(\rho)^{\otimes M}$,  followed by a generally joint POVM used to infer the pattern. The classification error probability is consequently,
\begin{equation}
p_{\text{err}}(\mathcal{B}) \defeq  \sum_{\bs{i} \neq \bs{i}^{\prime}} \pi_{\bs{i}} \text{Tr}\left[ {\Pi}_{\bs{i}^{\prime}} \rho_{\bs{i}}^{\otimes M}\right] . \label{eq:p_err_block}
\end{equation}
This protocol may be assisted by ancillary quantum systems that are free from channel interaction (often referred to as reference or idler modes), allowing for the use of entangled probe states which can lead to critical performance enhancements.
The optimal block protocol is determined via a double optimisation over the input state, and the discrimination POVM. It is clear that the general adaptive protocol inherits all simpler block protocols and therefore the optimal error rate $p_{\text{err}}^{\text{opt}}$ always satisfies,
\begin{equation}
p_{\text{err}}^{\text{opt}}\left( \mathcal{P}\right) \defeq \underset{\mathcal{P}}{\inf} \left[ p_{\text{err}}(\mathcal{P})\right] \leq   \underset{\mathcal{B}}{\inf} \left[ p_{\text{err}}(\mathcal{B})\right] \eqdef p_{\text{err}}^{\text{opt}}\left( \mathcal{B}\right)
\end{equation}

\subsection{Teleportation Stretching}
Restriction to the space of block/block-assisted protocols $\mathcal{B}$ simplifies the investigation of optimal error bounds for multi-channel discrimination, at the expense of losing generality. However, there is a prominent class of channels for which the optimal block-assisted and general adaptive strategies coincide: Teleportation-covariant channels. A quantum channel admits telecovariance if for any input state $\rho$ and pair of teleportation unitaries $\{U,V\}$, it satisfies,
\begin{equation}
\mathcal{E}(U\rho U^{\dagger}) = V\mathcal{E}(\rho)V^{\dagger}.
\end{equation}
This condition means that we can \textit{simulate} the channel $\mathcal{E}$ via a programmable teleportation protocol $\mathcal{T}$ using its Choi-state $\rho_{\mathcal{E}}$ as the resource (or program) state,
\begin{equation}
\mathcal{E}(\rho) = \mathcal{T}(\rho \otimes \rho_{\mathcal{E}}).
\end{equation}
The simulation of multi-channels relies on the joint-telecovariance of a collection of channels i.e.\ $\mathcal{E}_{\bs{i}}^m$ can be simulated using the same teleportation protocol, but with (potentially) $m$ different resource states,
\begin{equation}
{\rho}_{\mathcal{E}_{\bs{i}}^m} \defeq \rho_{\mathcal{E}_{i_1}} \otimes \cdots \otimes \rho_{\mathcal{E}_{i_m}} = \bigotimes_{k=1}^m \rho_{\mathcal{E}_{\bs{i}_k}}.
\end{equation}
Teleportation simulation refers to just one programmable protocol that can be used for channel simulation \cite{ConvBanchi, FLQCD}, but is particularly convenient for telecovariant channels since their program states take such an expedient form.\par
Bosonic quantum channels require a particularly careful treatment. In this case, the Braunstein-Kimble teleportation protocol is invoked and the teleportation unitaries take the form of phase-space displacement operators \cite{BKtele,PLOB}. Since the Choi-states of bosonic channels are energy unbounded, the simulation is asymptotic. Then these Choi states are actually asymptotic quasi-Choi matrices, defined as the sequence of finite-energy Choi approximations in the limit of infinite squeezing, 
\begin{equation}
\rho_{\mathcal{E}} \defeq \lim_{a \rightarrow \infty}\{ \rho_{\mathcal{E}}^{a}\} =  \lim_{a \rightarrow \infty} (\mathcal{E}\otimes \mathcal{I})(\Phi_{a}).
\end{equation}
Here $\Phi_{a}$ is a finite-energy TMSV state with $a = \bar{n}_S + \frac{1}{2}$, $\bar{n}_S$ is the mean photon number of the state, and $\mathcal{I}$ is the identity channel. This asymptotic treatment extends to all possible functionals taken on the quasi-Choi matrices, i.e.\ any $n$-state functional $f$ of asymptotic Choi-states is computed in the infinite squeezing limit over Choi-sequences $(\{\rho_{\mathcal{E}_1}^a\}, \ldots , \{\rho_{\mathcal{E}_n}^a\})$,
\begin{equation}
f(\rho_{\mathcal{E}_1}, \ldots, \rho_{\mathcal{E}_n}) = \lim_{a \rightarrow \infty} f( \rho_{\mathcal{E}_1}^a, \ldots, \rho_{\mathcal{E}_n}^a ). \label{eq:asymp}
\end{equation}
which is implicitly utilised throughout this paper. 
\par
The utility of telecovariance in the context of discrimination will now become clear; given a quantum comb $\mathcal{P}$, one may replace the channel pattern slots with their respective multi-channel teleportation simulations $\mathcal{E}_{\bs{i}}^m(\rho) = \mathcal{T}(\rho\otimes \rho_{\mathcal{E}_{\bs{i}}^{m}})$ (see Fig.~\ref{fig:adaptiveprotocol}(a)). Since teleportation is an LOCC protocol it can be summarised via a QO which is inherited by the subsequent adaptive operation $\Lambda_k, \> (k =1,...,M)$ that takes place at each pattern instance. This allows for the $M$ channel resource states $\rho_{\mathcal{E}_{\bs{i}}^m}^{\otimes M}$ to be \textit{stretched} back in time, outside of the adaptive operations at each round of pattern interaction. Furthermore, with all resource states stretched outside of the adaptive operations, these can be collapsed into a single trace-preserving QO, $\tilde{\Lambda}$. \par
Remarkably, teleportation stretching simplifies the quantum comb into the generation of $M$-resource states, followed by $\tilde{\Lambda}$ and a collective POVM, $\Pi_{\bs{i}}$. The output state prior to measurement can be expressed as,
\begin{equation}
\rho_{\bs{i}}^M = \tilde{\Lambda}( \rho_{\mathcal{E}_{\bs{i}}^m}^{\otimes M}).
\end{equation}
In the case of finite-dimensional systems, thanks to Naimark's dilation theorem any POVM can be represented as a quantum channel followed by a projective measurement \cite{HelstromNaimark, Paris2012}. This means that  $\tilde{\Lambda}$ can be further absorbed into the joint POVM $\tilde{\Pi}_{\bs{i}} = \Pi_{\bs{i}} \circ \tilde{\Lambda}$, completing the reduction from general adaptive protocol, to a block-assisted protocol $\mathcal{P} \rightarrow \mathcal{B}$. Then, the error rate of the optimal general adaptive protocol is equal to the error rate of the optimal block assisted protocol $p_{\text{err}}^{\text{opt}}\left( \mathcal{P}\right) = p_{\text{err}}^{\text{opt}} ( \mathcal{B} ,\tilde{\Lambda} ) = p_{\text{err}}^{\text{opt}}\left( \mathcal{B} \right)$, where $(\mathcal{B} ,\tilde{\Lambda} ) $ is a block assisted protocol aided by a pre-measurement adaptive operation. Indeed, the optimal probability of misclassification for a generic image space $\mathcal{U} = \{\pi_{\bs{i}}, \mathcal{E}_{\bs{i}}^m\}$ is given by
\begin{equation}
p_{\text{err}}^{\text{opt}}(\mathcal{P})  = p_{\text{err}}^{\text{opt}}(\mathcal{B}) = \inf_{\tilde{\Pi}_{\bs{i}^{\prime}}} \left[ \sum_{\bs{i} \neq \bs{i}^{\prime}} \pi_{\bs{i}} \text{Tr}\left[ \tilde{\Pi}_{\bs{i}^\prime} \rho_{\mathcal{E}_{\bs{i}}^m}^{\otimes M}\right] \right]. \label{eq:ult_perf}
\end{equation}
Deriving a lower bound for $p_{\text{err}}^{\text{opt}}(\mathcal{B})$ immediately implies a fundamental limit. \par
The same simplification may not apply for infinite dimensional systems, thus we provide a more general treatment by exploiting properties of the fidelity. Crucially, since trace preserving operations $\tilde{\Lambda}$ cannot increase the distance between two quantum states, one can write the data-processing inequality,
\begin{align}
F(\rho_{\bs{i}}^{M}, \rho_{\bs{i}^\prime}^{M}) &= F \left( \Lambda \left( \rho_{\mathcal{E}_{\bs{i}}^m}^{\otimes M}\right),  \Lambda \left( \rho_{\mathcal{E}_{\bs{i}^\prime}^m}^{\otimes M} \right)\right),\\
&\geq F \left( \rho_{\mathcal{E}_{\bs{i}}^m}^{\otimes M} , \rho_{\mathcal{E}_{\bs{i}^\prime}^m}^{\otimes M} \right), \label{eq:lbound}
\end{align}
where the Bures fidelity $F(\rho,\sigma)  = \| \sqrt{\rho} \sqrt{\sigma} \|_1 = \text{Tr} \left[ \sqrt{\sqrt{\rho} \sigma \sqrt{\rho}}\right]$ is computed asymptotically (as in Eq.~(\ref{eq:asymp}). It then follows that this lower bound must hold for any possible general adaptive protocol $\mathcal{P}$, and all $\bs{i},\bs{i}^{\prime} \in \mathcal{U}$,
\begin{equation}
F_{\text{opt}} \defeq \underset{\mathcal{P}}{\text{inf}} \left[ F(\rho_{\bs{i}}^M, \rho_{\bs{i}^\prime}^M)\right] \geq  F^M ( \rho_{\mathcal{E}_{\bs{i}}^m}, \rho_{\mathcal{E}_{{\bs{i}}^\prime}^m} ). \label{eq:supP}
\end{equation}
It clear that this fidelity lower bound is achievable. Consider the block-assisted protocol $\mathcal{B}$, shown in Fig.~\ref{fig:adaptiveprotocol}(c), that probes each channel $i_k \in \bs{i}$ in the pattern via $M$-copy TMSV states $\Phi_{a}^{\otimes M}$, while retaining idler modes for each channel. In the limit of infinite squeezing $(a\rightarrow \infty)$ this results in the output state $\rho_{\mathcal{E}_{\bs{i}^\prime}^m}^{\otimes M}$ prior to collective measurement, satisfying equality in Eq.~(\ref{eq:supP}). Thus the ultimate performance is achieved via $\mathcal{B}$ so that $F_{\text{opt}} =  F^M ( \rho_{\mathcal{E}_{\bs{i}}^m}, \rho_{\mathcal{E}_{{\bs{i}}^\prime}^m} )$. Substituting this result into the fidelity based state lower bound from \cite{LBMont}, we gather
\begin{equation}
p_{\text{err}}^{\text{opt}}(\mathcal{P}) \geq  \frac{1}{2} \sum_{\bs{i} \neq \bs{i}^{\prime}} \pi_{\bs{i}} \pi_{\bs{i}^\prime} F^{2M}(\rho_{\mathcal{E}_{\bs{i}}^m} , \rho_{\mathcal{E}_{\bs{i}^{\prime}}^m}) \label{eq:LB}.
\end{equation}
This represents a readily computable ultimate lower bound on the optimal error probability of classifying binary, jointly telecovariant channel patterns.\par

\subsection{Ultimate Limits and Quantum Advantage}
In order to know when quantum protocols are advantageous over classical protocols, it is necessary to derive an upper bound on $p_{\text{err}}^{\text{opt}}(\mathcal{P})$ that encompasses the use of quantum resources.
An expedient quantum upper bound can be derived via a restriction to local measurements on the optimal block-assisted protocol $\mathcal{B}^{\text{loc}}$, such that ${\Pi}_{\bs{j}} = \bigotimes_{k=1}^m \Pi_{j_{k}}$. The discriminatory POVMs are now completely separable, and as such the global classification error is defined as the compounded error of individual channel discriminations, $p_{\text{err}}^{\text{opt}}(\mathcal{P})  \leq p_{\text{err}}^{\text{opt}}(\mathcal{B}^{\text{loc}}) = 1 - (1 - p_{\text{err}}^{\text{pixel}})^m $ achieved by local Helstrom measurements. Using the analyses from \cite{PirBD, PatternRecog}, considering $m$-length, uniformly distributed channel patterns probed by $M$-copy infinitely squeezed TMSV states we can write
\begin{equation}
p_{\text{err}}^{\text{opt}}(\mathcal{P})  \leq 1 - \Big[ 1 - \frac{1}{2} {F^M(\rho_{\mathcal{E}_B}, \rho_{\mathcal{E}_T}}) \Big]^m. \label{eq:LocalM}
\end{equation}
For non-uniform channel distributions, it may instead be useful to utilise the upper bound,
\begin{align}
p_{\text{err}}^{\text{opt}}(\mathcal{P}) \leq \sum_{\bs{i} \neq \bs{i}^{\prime}} \sqrt{\pi_{\bs{i}} \pi_{\bs{i}^\prime}} F^{M}  ( \rho_{\mathcal{E}_{\bs{i}}^m}, \rho_{\mathcal{E}_{\bs{i}^\prime}^m} ), \label{eq:JointUB}
\end{align}
based on Pretty Good Measurements (PGMs) \cite{UBPGM}, however Eq.~(\ref{eq:LocalM}) is always superior for uniformly distributed patterns.\par
Furthermore, it is important to characterise a lower bound on the optimal performance of protocols limited to classical resources. The best known classical strategy is an unassisted block-protocol with $p_{\text{err}}^{\text{opt}}(\mathcal{B}^{\text{cl}})$ such that $\mathcal{B}^{\text{cl}} \subset \mathcal{B}$. Such protocols are restricted to the use of coherent, classical input states, but have access to global measurements. Fortunately, since the thermal images are modelled using phase-independent channels with identical transmissivities, the application of operations that induce displacement or phase shifts on the input states have no effect on the outputs. Also, the optimal classical input state is pure; hence, the optimal classical strategy may simply make use of $M$-copy vacuum states $\ket{0}^{\otimes M}$ in order to probe each channel, followed by a joint measurement. Substituting $\mathcal{E}_{\bs{i}}^m (\ket{0}_{1} \otimes \ldots \otimes \ket{0}_m)^{\otimes M}$ into Eq.~(\ref{eq:LB}) the classical lower bound is derived.  \par

It is now possible to qualify, and quantify quantum advantage. Guaranteed quantum advantage defines the point at which a quantum-enhanced protocol outperforms the optimal classical strategy for a given discrimination task. In the case of uniformly distributed channel patterns, using the knowledge that block-assisted protocols limited to local measurements upper bound the optimal quantum performance, we can say that quantum advantage is guaranteed when
\begin{equation}
p_{\text{err}}^{\text{opt}}(\mathcal{B}^{\text{loc}}) \leq p_{\text{err}}^{\text{opt}}(\mathcal{B}^{\text{cl}}).
\end{equation}
Since we cannot work directly with these quantities, advantage is only guaranteed whenever the lower bound of the classical strategy is outperformed by the upper bound of the quantum-enhanced strategy. Defining upper and lower bounds for classical (quantum) discrimination protocols,  $p_{\text{err}}^{\text{cl,U}}$ and $p_{\text{err}}^{\text{cl,L}}$ ($p_{\text{err}}^{\text{q,U}}$ and $p_{\text{err}}^{\text{q,L}}$) we define the \textit{minimum guaranteed advantage} (MGA) as the minimum performance enhancement achieved by a quantum strategy over the optimal classical one,
\begin{equation}
\Delta p_{\text{err}}^{\text{min}} \defeq p_{\text{err}}^{\text{cl,L}} - p_{\text{err}}^{\text{q,U}}. \label{eq:MGA}
\end{equation}
If $\Delta p_{\text{err}}^{\text{min}} > 0$, then quantum advantage is guaranteed. One can also define the \textit{maximum potential advantage} (MPA) as
\begin{equation}
\Delta p_{\text{err}}^{\text{max}} \defeq p_{\text{err}}^{\text{cl,L}} - p_{\text{err}}^{\text{q,L}}. \label{eq:MPA}
\end{equation}
This represents the maximum possible improvement quantum strategies can bring when the derived quantum lower bound is fundamental. It is also only a \textit{potential} advantage, and $\Delta p_{\text{err}}^{\text{max}}$ may be positive when $\Delta p_{\text{err}}^{\text{min}} \leq 0$ in which case advantage is not certified. \par
When determining conditions for advantage, it is also useful to devise a quantity that generally measures resource demands. To this end, we define the quantity 
\begin{equation}
\bar{M} \defeq \frac{M}{m},
\end{equation} as the \textit{relative probe copy number} of a given block protocol. This describes the number of probe copies required for a given discrimination protocol \textit{relative} to the dimension of the channel pattern in question. A large $\bar{M}$ occurs when $M \gg m$, and implies a high resource cost since each channel in the pattern requires many probings prior to measurement. Contrarily, a small value of $\bar{M}$ implies a very low resource cost relative to the size of the image. Determining a \textit{minimum} relative probe copy number defining the minimum $\bar{M}$ required to guarantee quantum advantage,
\begin{equation}
\bar{M}_{\text{adv}} = \argmin_{\bar{M}}\left( | \Delta p_{\text{err}}^{\text{min}} |\right) \label{eq:MinRelPrN},
\end{equation}
i.e.~$\bar{M}_{\text{adv}} = \bar{M}$ such that $\Delta p_{\text{err}}^{\text{min}} = 0$, which provides a very useful metric on the resource cost and feasibility of a discrimination protocol.\par
While this characterisation of advantage relies on the use of infinitely squeezed input states, realistic quantum protocols must make use of finite-energy inputs. Finite-squeezing effects can be taken into account via the consideration of finite-energy TMSV states as probe inputs, rendering output states to be  Choi-state approximations. The introduction of any two-mode squeezing immediately introduces stronger distinguishability over the classical vacuum states, and considerable advantage can be attained for realistic energy resources (see Appendix \ref{sec:FEFids} for more details).

\section{Exact Classification of Thermal Images\label{sec:ExactClass}}
\begin{figure}[t!]
\hspace{1.2cm}(a)\\
\includegraphics[width=0.8\linewidth]{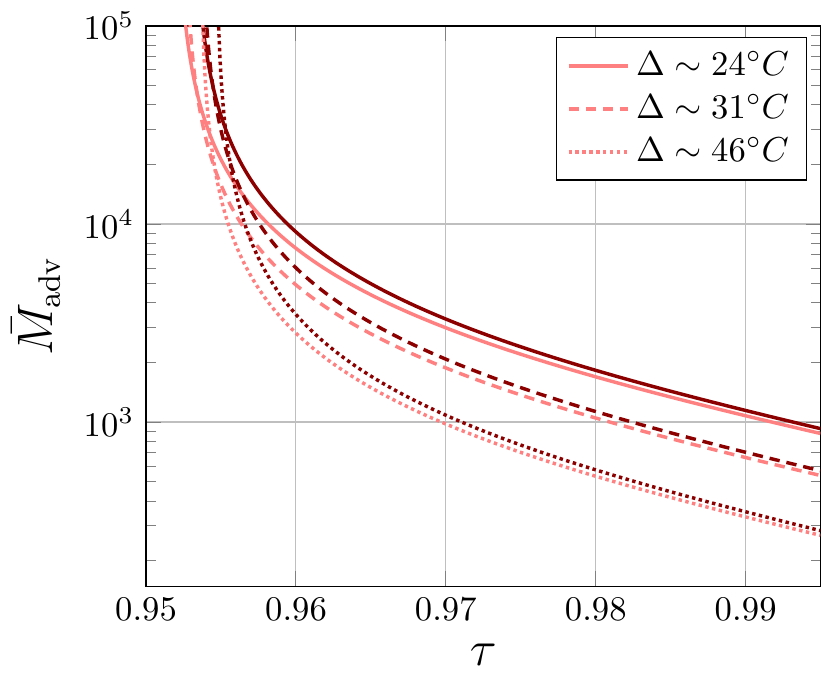} \\
\hspace{1.2cm}(b)\\ \hspace{1cm}\\
\includegraphics[width=0.85\linewidth]{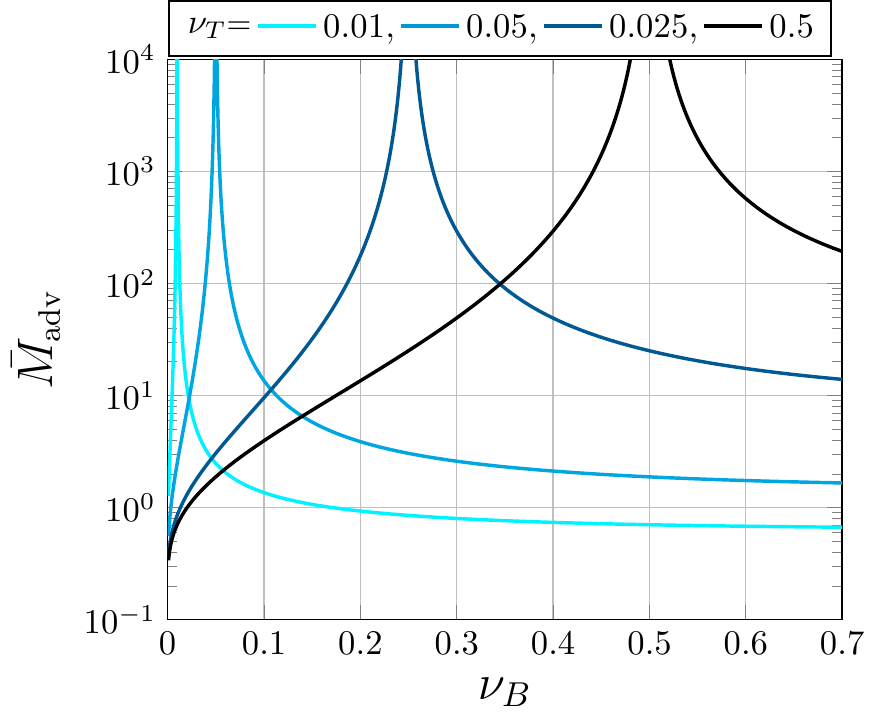}
\caption{Minimum relative probe copy number $\bar{M}_{\text{adv}}$ required for guaranteed quantum advantage, with respect to: (a) Thermal-loss channels under low-loss and different thermal noise parameters, and (b) additive noise channels under different noise conditions. In (a) we model images using thermal target/background channels with pixel temperature variance $\Delta^{\circ}C$ (in a microwave setting with wavelength $\sim 1\text{mm}$, see Appendix \ref{sec:Temps}). The darker plots describe images with this variance at higher temperatures within $[0,50]^{\circ}C$, whilst the brighter plots describe inherently colder environments within $[-10,40]^{\circ}C$ . In (b), the background channel noise $\nu_B$ is varied against the fixed target noise for a number of examples.
}
\label{fig:GAdv}
\end{figure}
\begin{figure}[t!]
\hspace{0.5cm} (a) Thermal-Loss, $m=9$ \hspace{0.2cm} (b) Additive Noise, $m=9$\hspace{0.5cm}  \\
\includegraphics[width=0.515\linewidth]{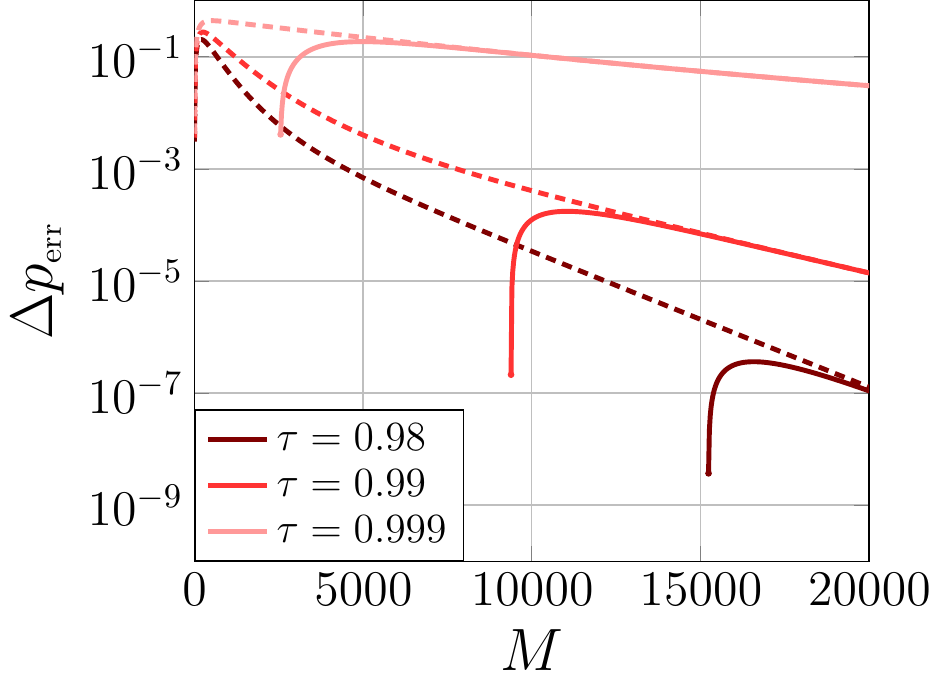}
\includegraphics[width=0.465\linewidth]{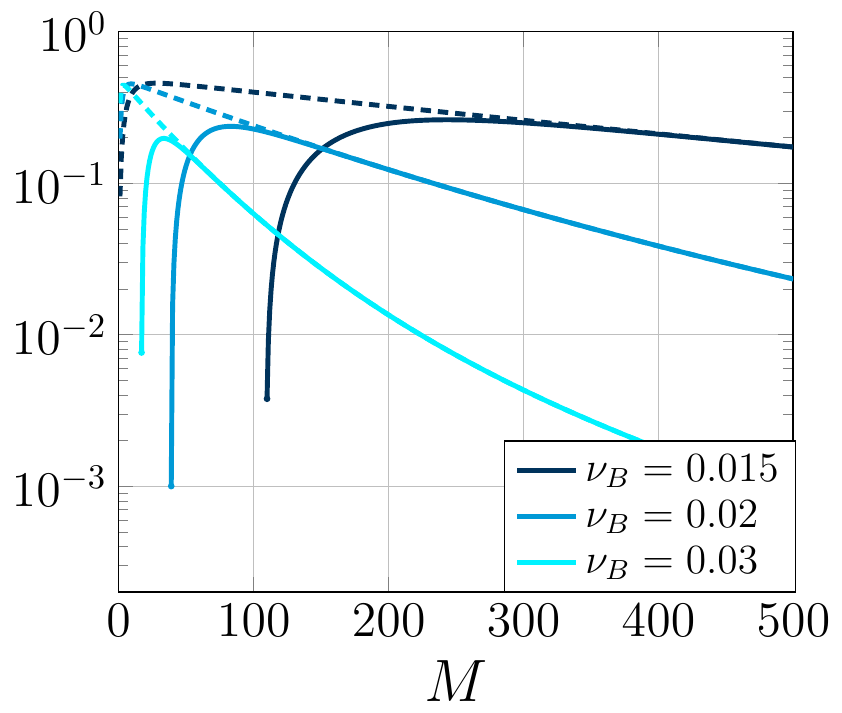}\\
\hspace{0.5cm} (c) Thermal-Loss, $m=9$ \hspace{0.2cm} (d) Additive Noise, $m=50$\hspace{0.5cm}  \\
\includegraphics[width=0.5125\linewidth]{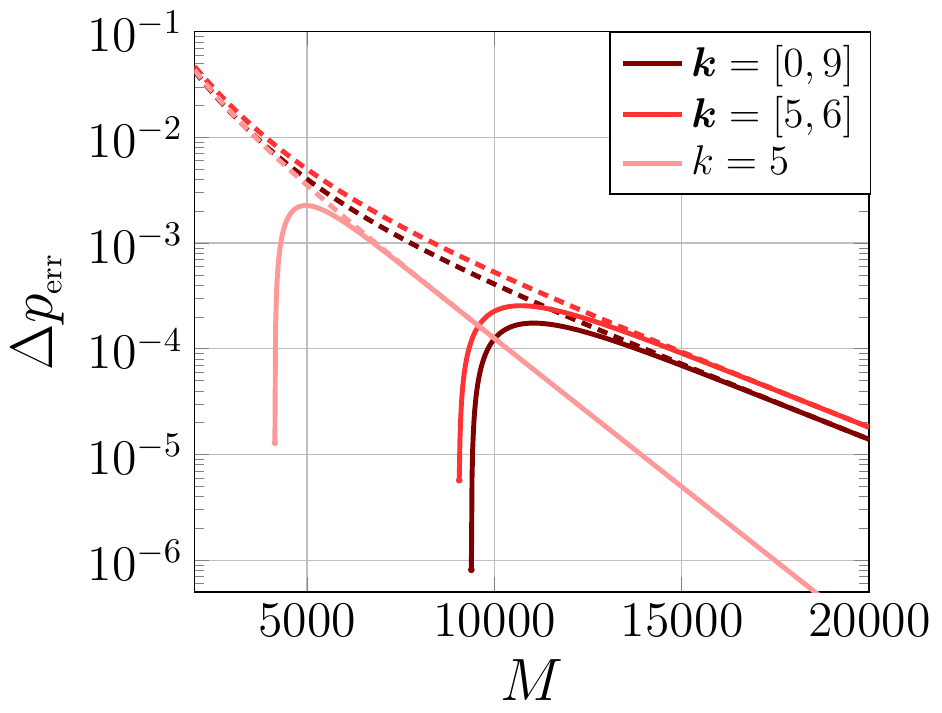}
\includegraphics[width=0.4725\linewidth]{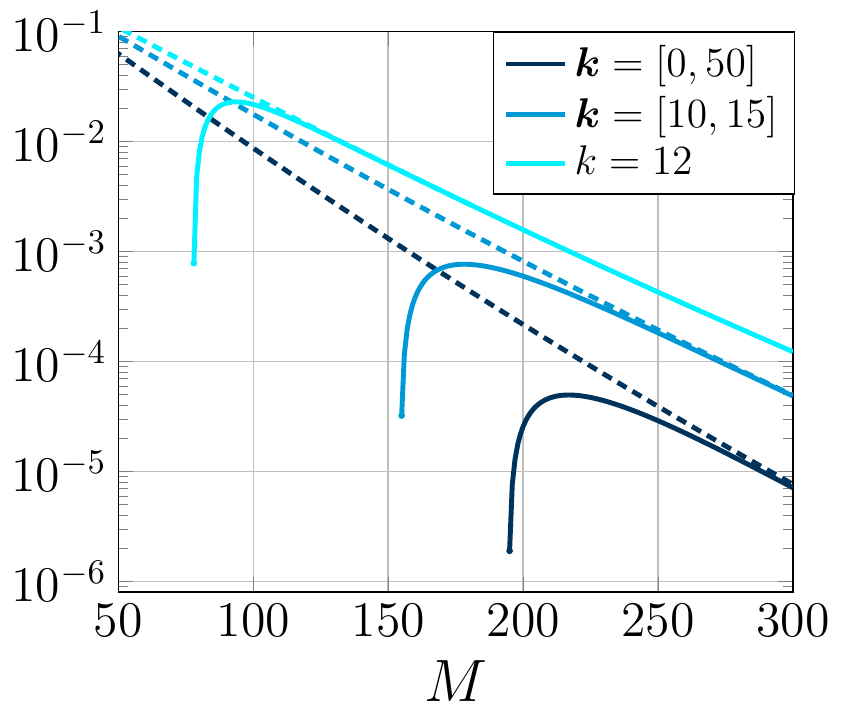}
\caption{
Panels (a) and (b) consider ultimate bounds on thermal image classification based on a uniform distribution of $m$ target channels, $\bs{i} \in \mathcal{U}_m$. Panel (a) considers $m=9$ pixel images such that each channel is a thermal loss channel under conditions $\{\epsilon_B,\epsilon_T\} = \{18.5,20.2\}$ and different values of transmissivity $\tau$ in the low-loss regime. In Panel (b) this is shown for additive-noise channel patterns $ \nu_T  = 0.01$ for $m=9$ pixels and different values of background noisy channels $\nu_B > \nu_T$. Panels (c) and (d) focus on $k$-BCPF image spaces ${\bs{i}}\in \bigcup_{k\in{\bs{k}}} \mathcal{U}_{{\textsf{\tiny CPF}}}^{k}$. Panel (c) considers $m=9$ pixel images such that each channel is a thermal loss channel under conditions $\{\tau,\epsilon_B,\epsilon_T\} = \{0.99,18.5,20.2\}$. In Panel (d) this is shown for additive-noise channel patterns $\{\nu_B,\nu_T\} = \{0.02,0.01\}$ for $m=50$ pixels.
}
\label{fig:ThermalBcodes}
\end{figure}
\subsection{Uniform Image Spaces \label{sec:Barcodes}}
Consider the $m$-pixel uniform image space $\mathcal{U}_m$ such that each pixel is equally likely to be a target or background environment. Thanks to the multiplicativity of the fidelity, the multimodal fidelity of the multi-channel Choi states can be simplified to 
\begin{equation}
F({\rho}_{\mathcal{E}_{\bs{i}}^m}, {\rho}_{\mathcal{E}_{{\bs{i}^\prime}}^m}) = F({\rho}_{\mathcal{E}_{B}}, {\rho}_{\mathcal{E}_{T}} )^{\text{hamming}(\bs{i},{\bs{i}^\prime})}.
\end{equation}
Employing the teleportation stretching lower bound Eq.~(\ref{eq:LB}), and exploiting properties of the Hamming distance \cite{PatternRecog} one can then derive an ultimate lower bound in terms of fidelity between single copy Choi states. In conjunction with the $\mathcal{B}^{\text{loc}}$ upper bound Eq.~(\ref{eq:LocalM}), the optimal error probability is bounded via,
\begin{align}
& p_\text{err}^{\text{opt}} \geq \frac{[ F^{2M}(\rho_{\mathcal{E}_{T}}, \rho_{\mathcal{E}_{W}} )+ 1]^m - 1}{2^{m+1}}, \label{eq:LB_Bcode}\\
&p_{\text{err}}^{\text{opt}} \leq 1 - \Big[ 1 - \frac{1}{2} {F^M(\rho_{\mathcal{E}_B}, \rho_{\mathcal{E}_T}}) \Big]^m \label{eq:UniUB}.
\end{align}
Using Bernoulli's inequality $(1+x)^n \geq 1+nx$ in conjunction with Eqs.~(\ref{eq:UniUB}) and (\ref{eq:LB_Bcode}) these bounds can be more succinctly presented,
\begin{equation}
\frac{m}{2^{m+1}} F^{2M} \leq p_\text{err}^{\text{opt}} \leq  \frac{m}{2} F^{M}.
\end{equation}
It is now easy to qualify and quantify guaranteed advantage for a uniformly distributed $m$-pixel classification problem,
\begin{equation}
 \Delta p_{\text{err}}^{\text{min}} > 0 \implies  F_{\text{cl}}^{2M} > 2^m F_{\text{q}}^{M} \label{eq:UniAdv}.
\end{equation}
Using Eq.~(\ref{eq:UniAdv}) and $M = \bar{M} m$, it can be shown that the minimum relative probe copy number required to guarantee quantum advantage is given by 
\begin{equation}
\bar{M}_{\text{adv}} = \frac{\log 2}{2\log({F_{\text{cl}}}) - \log({F_{\text{q}}})}, \label{eq:MinPCN}
\end{equation}
which is analysed in Fig.~\ref{fig:GAdv}. Thanks to the compactness of the additive-noise Choi state fidelities, we can write
\begin{align}
{\bar{M}}_{\text{adv}} &= \frac{\log2}{\log{(1+\Delta_q)} - 2\log(1+\Delta_{\text{cl}}) },\\
\Delta_\text{q}  &\defeq \frac{(\sqrt{\nu_B}  - \sqrt{\nu_T})^2}{2\sqrt{\nu_B \nu_T}}, \\
\Delta_\text{cl} &\defeq \sqrt{(\nu_T+1)(\nu_B+1)} - \sqrt{\nu_B \nu_T} - 1,
\end{align}
while the thermal-loss/amplifier result is too long to display, but can of course be treated via Eq.~(\ref{eq:MinPCN}). Using any $\bar{M} > \bar{M}_{\text{adv}}$ guarantees advantage.\par
Characterising $\bar{M}_{\text{adv}}$ allows us to predict the necessary resource demands of a given thermal imaging setting for general $m$ and $M$. In Fig.~\ref{fig:GAdv}(a) we report resource costs for pattern discrimination at different pixel temperature variances $\Delta^\circ C$. From these results, it can be seen that discrimination is (unsurprisingly) easier at lower temperatures, since colder temperatures will endue less decoherence onto incident probes. Furthermore, advantage has a much greater dependence on the pixel variance $\Delta$, than the specific temperatures. Nonetheless, it is clear that quantum advantage is only viable in the very low-loss limit, since $\bar{M}_{\text{adv}} \rightarrow \infty$ for $\tau \sim 0.95$. \par
 As $\tau \rightarrow 1$ the prospect of advantage becomes increasingly likely, and when additive noise channels are considered (such that $\tau = 1$), guaranteed advantage is readily attainable for a variety of channel pattern conditions. Advantage is achieved at a low resource cost when \textit{either} the target or background channel are of low additive noise (seen in Fig.~\ref{fig:GAdv}(b)). In some very low noise settings $\bar{M}_{\text{adv}} \approx 1 $, meaning that the necessary number of probe copies to guarantee advantage scales with the dimension of the pattern $M \sim m$. If $\bar{M}_{\text{adv}} < 1$ then $M \lesssim m$ and discrimination is extremely cost effective via quantum resources. \par
Further results are shown in Fig.~\ref{fig:ThermalBcodes} for $m=9$ pixel images, where we plot the MGA and MPA $\Delta p_{\text{err}}$ with respect to probe number, for both channels and channel properties. The initial point at which the MGA is plotted is the first value of $M$ for which the advantage inequality is satisfied, and determines the minimum number of probes required to guarantee quantum advantage.

\subsection{Channel Position Finding}
As discussed in Section \ref{sec:TeleCov}, one may consider non-uniform image spaces such as those formulated using channel position finding. Obtaining \textit{a priori} knowledge of the image space can lead to dramatic improvements in discrimination performance, and is immensely useful for quantum illumination, reading and more. We begin with the concept of $k$-CPF, such that each $m$-pixel image in the space $\bs{i} \in \mathcal{U}_{\textsf{\tiny CPF}}^{k}$  contains exactly $k$-target channels. Then the image space dimension has been dramatically reduced from $2^m$ to $|{\mathcal{U}_{{\textsf{\tiny CPF}}}^k}| = {m\choose k}$ (where ${m\choose k}$ is the binomial coefficient). Deriving error bounds requires us to find a solution to 
\begin{equation}
D_m^k = \frac{1}{{m\choose k}} {\sum_{ (\bs{i}\neq\bs{i}^\prime) \in{\mathcal{U}_{{\textsf{\tiny CPF}}}^k} }} f^{\text{hamming}(\bs{i},\bs{i}^\prime)}.
\end{equation}
Once more exploiting properties of the Hamming distance and a straightforward counting argument (see Appendix \ref{sec:CPF}), we find a closed formula in terms of the standard hypergeometric function,
\begin{equation}
D_m^k [f] =  {{}_2 F_1}(-k,k-n,1,f^2) - 1 \label{eq:Strictk}.
\end{equation}
Substituting into the bounds from Eqs.~(\ref{eq:LB}) and (\ref{eq:JointUB}), and assuming each pattern in the image space occurs with equal probability, $\pi_{\bs{i}} = {m \choose k}^{-1}$, then 
\begin{align}
\frac{D_m^k \left[ F^{2M} \right]}{2{m\choose k}} \leq p_{\text{err}}^{\text{opt}} \leq D_m^k \left[ F^{M} \right].
\end{align}
One may more generally consider the case that there exists an image space $\mathcal{U}_{\textsf{\tiny CPF}}^{\bs{k}} = \bigcup_{k\in\bs{k}} \mathcal{U}_{\textsf{\tiny CPF}}^k$ such that $\bs{k}$ contains all possible numbers of target channels in any image in the space i.e.\  we have both an upper and lower bound on the number of targets, $\bs{k} = \{ k_{\text{min}}, \ldots, k_{\text{max}}\}$, or even just a suspected set of $T$ target numbers $\bs{k} = \{k_1, \ldots, k_T\}$. We call this $k$-Bounded CPF ($k$-BCPF) as we bound the number of targets in any possible image, generalising the previous results. In order to derive error probability bounds we must solve,
\begin{equation}
D_m^{\bs{k}}(f) = \sum_{ (\bs{i}\neq\bs{i}^\prime) \in{\mathcal{U}_{\textsf{\tiny CPF}}^{\bs{k}}} } f^{\text{hamming}(\bs{i},\bs{i}^\prime)} . \label{eq:Boundk}
\end{equation}
By splitting this quantity into its diagonal (patterns drawn from likewise image spaces) and off diagonal (patterns drawn from unequal image spaces) sums, then it can be solved using similar counting arguments as before. Indeed, defining the following functionals, 
\begin{equation}
D_m^{\bs{k}} (f) = \sum_{j \in \bs{k} } {D_{m}^j(f)}+  \sum_{ i \in \bs{k} } \sum_{l \neq i} {\tilde{D}_{m}^{i,l}(f)} \label{eq:Boundk2}.
\end{equation}
where $D_m^j$ is given in Eq.~(\ref{eq:Strictk}) and 
\begin{equation}
\tilde{D}_m^{k,l}(f) = {m\choose l} {l\choose k} f^{k-l} {}_2 F_1 (-k, l-m, l-k+1, f^2).
\end{equation}
Then one can write upper and lower bounds for $k$-BCPF using Eqs.~(\ref{eq:Strictk}) and (\ref{eq:Boundk2}), in conjunction with uniform weights,
$
\tilde{\pi}_{\bs{k}} = \left[ \sum_{i \in \bs{k} }  {m\choose i} \right]^{-1},
$
\begin{align}
\frac{\tilde{\pi}_{\bs{k}}^2}{2}  D_m^{\bs{k}} \left[ F^{2M} \right] \leq p_{\text{err}}^{\text{opt}} &\leq {\tilde{\pi}_{\bs{k}}}  D_m^{\bs{k}} \left[ F^{M} \right].
\end{align}
Fig.~\ref{fig:ThermalBcodes}(c) and (d) depict examples of these bounds for $m=9$ and $m=50$ pixel images. These results emphasise that precise knowledge of a target channel distribution has a significant role on the amount of advantage attainable through quantum sources. Ambiguity over the target distribution (even by one pixel) can remove any enhancement obtained over the discrimination of uniform channel patterns.

\subsection{Distinguishability and Decoherence}
In all the settings studied thus far, it is immediately obvious that the number of probes required to prove advantage for thermal-loss channels is very high, and increasingly so for lower transmissivity. Protection of the shared probe-idler entanglement is of course critical to improve the distinguishability of output states. When considering additive-noise channels, we isolate the role of thermal noise on the ultimate performance limits. While noise on its own is capable of degrading the quality of this entanglement, at sufficiently low-levels the quantum advantage is expedient and very effective. But when loss is combined with this noise, this endues a rapid decay in the shared probe-idler entanglement, as these decohering effects cooperate to reduce the distinguishability of output states.\par 
It is therefore clear that idler-assisted discrimination protocols will prove quantum advantage only in the low-loss limit. At high loss, the ultimate and classical lower bounds quickly coincide, removing any benefit from utilising quantum sources. Such properties can be satisfied in the short-range imaging of thermal environments.

\section{Quantum-Enhanced Pattern Recognition \label{sec:SLearn}}
When dealing with a small number of pixels, the bounds in the previous section emphasise that there exist quantum advantageous protocols for image classification, achievable with realistic channel environments. However as the number of pixels increases, performing this exact classification method leads to a high resource demand in order to achieve low error probability, as the discriminator attempts to classify a single pattern from an exponentially growing set of possible patterns. Instead, we can make use of machine learning techniques, commonplace in image classification problems, and enhance them with quantum sensors.
\begin{figure*}[t!]
\hspace{1cm} (a) \hspace{8cm} (b)\\
\includegraphics[width=0.5\linewidth]{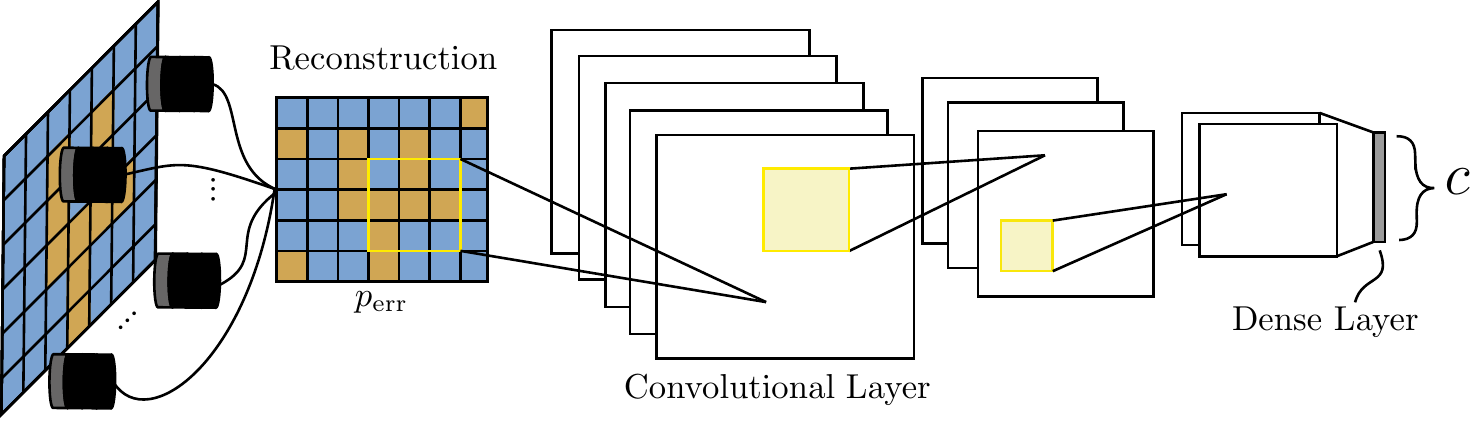}
\hspace{1cm} \includegraphics[width=0.35\linewidth]{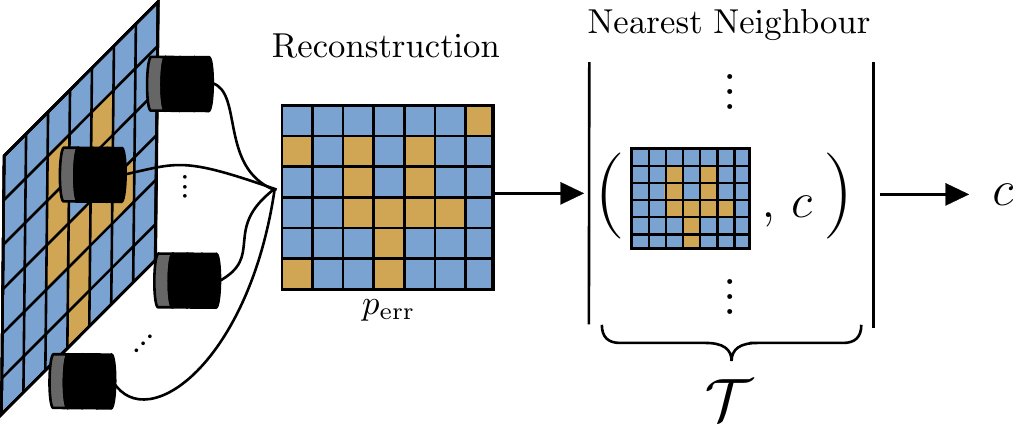}
\caption{Image classification via supervised machine learning post processing. Panel (a) depicts a (potentially quantum) sensor probing a thermal image, which is then classified via a convolutional neural network (CNN) which has been trained on similar data. The result of the classification is encoded in the final layer of the network. Panel (b) illustrates the use of a nearest neighbour classifier, which infers the class of a detected image by locating its closest (in terms of some distance measure) image in a training set $\mathcal{T}$.}
\label{fig:Classifiers}
\end{figure*}
\subsection{Nearest Neighbour Classifier}
Image classification can reframed as follows: Suppose there exists a space of all possible $m$-pixel images, such that the probability of obtaining an image $\bs{i}$ is $p_{\bs{i}}$. Each image has an associated class (or label) $c$, connected via a conditional probability distribution $P(c|\bs{i})$. Since this probability distribution is generally unknown, we design a classifier $\tilde{c}(\bs{i})$ that approximates its behaviour with a probability of error $E$. The smaller the error probability, the closer the approximate distribution is to reality. That is, an optimal classifier satisfies,
\begin{equation}
\tilde{c}_{\text{B}}(\bs{i}) = \underset{c}{\text{argmax}}[P(c|\bs{i})]
\end{equation}
and is known as the Bayes classifier, such that 
\begin{equation}
E_{\text{B}} = 1 - \mathbb{E}_{\bs{i}}[P(\tilde{c}_{\text{B}}(\bs{i}) | \bs{i})]
\end{equation}
is its associated Bayes risk, which represents the minimum possible error probability of such a classifier. If a classification problem involves images each of which belong to one of $N$ distinct classes, then the $E_{\text{B}} = 0$, since there exists an optimal classifier than can perfectly distinguish between all images.\par
However as stated, $P(c|\bs{i})$ is generally unknown, and its therefore necessary to provide an approximation to this decision rule. This is often achieved with the assistance of a training set $\mathcal{T} = \{c_k ; \bs{i}_k\}_k^\text{tr}$ of size $T = |\mathcal{T}|$ which contains \text{pre}-labelled images. The decision rule can then be modelled with a conditional probability distribution $P(c|\bs{i}, \mathcal{T})$ such that the classification of an image $\bs{i}$ depends on expertise drawn from $\mathcal{T}$.\par
A well studied option is the Nearest Neighbour (NN) classifier, characterised by the function,
\begin{equation}
\tilde{c}_{\text{NN}}(\bs{i}) = c \Big( \underset{{\bs{i}^\prime} \in \mathcal{T}}{\text{argmin}}[ d(\bs{i}, {\bs{i}^\prime})] \Big)
\end{equation}
such that $\tilde{c}_{\text{NN}}$ retrieves the label of the closest image ${\bs{i}^\prime} \in \mathcal{T}$ to the sampled image $\bs{i}$ that we wish to classify. ``Closeness" in this context is defined by some appropriate distance measure $d(\bs{i},{\bs{i}^\prime})$ natural to the data in question. Using the NN classifier, and given a training set $\mathcal{T}$ of size $T$, one can write the expected classification error as
\begin{equation}
E_{\text{NN}}^T = \underset{\bs{i},\mathcal{T}}{\mathbb{E}} \left[ \sum_{c\neq\tilde{c}_{\text{NN}}} P(c|\bs{i})P(\tilde{c}_{\text{NN}}|\bs{i},\mathcal{T})\right] \label{eq:error}
\end{equation}
where we have recognised that the dependence on $\mathcal{T}$ has made the classifier itself a random variable. This quantity is the expected probability of misclassification, taken over image space and training sets of dimension $T$. \par
It is now important to note that in the limit of $T \rightarrow \infty$, provided that all $N$ classes are distinct and $N \ll 2^{m}$, the classifier tends towards optimality and the Bayes risk (as shown in \cite{PatternRecog}). As the training set increases in size, the likelihood of a sampled state $\bs{i}^\prime$ that we wish to classify also being contained in $\mathcal{T}$ tends towards unity. That is the probability of locating the closest image $\bs{i}^\prime$ to $\bs{i}$ becomes
\begin{equation}
\lim_{T\rightarrow \infty} P_{\text{NN}}(\bs{i}^\prime | \bs{i},\mathcal{T})  = \delta_{\bs{i}^\prime, \bs{i}}
\end{equation}
and in turn,
\begin{equation}
\lim_{T\rightarrow \infty} E_{\text{NN}} = E_{\text{B}} = 0.
\end{equation}
Hence, in the limit of infinite training set size the NN classifier achieves the Bayes rate. In reality we are unable to make use of infinite training sets, but it is still possible to numerically investigate finite-sample-risk and study the performance of such classifiers. To this end, one can make use of the result from \cite{Snapp},
\begin{equation}
\enn^T \approx \enn^\infty + \sum_{j=2}^{\infty} x_j T^{-j/m} \label{eq:Snapp}
\end{equation}
which we can treat as an ansatz to interpolate numerical results. Truncating this series at a sufficient value, the finite sample error rate can be readily approximated.\par

\subsection{Convolutional Neural Networks}
Given a set of training data as before $\mathcal{T} = \{c_k ; \bs{i}_k\}_k^\text{tr}$, a neural network is a universal approximating function $f_{\bs{\theta}}$ dependent on a set of variational parameters $\bs{\theta}$ that we wish to optimise in order to \textit{learn} a relationship between the input/outputs of the training set. The goal of training is to adjust $\bs{\theta}$ such that
\begin{equation}
f_{\bs{\theta}}(\bs{i}_k) \approx c_k, \> \forall k.
\end{equation}
Using a sufficiently large training set and variational parameter set, the network should be able to unveil (potentially highly non-linear) relationships between input images and their corresponding classification. \par
Feedforward neural networks form a particular architecture for this variational function, consistent of sequential layers of transformations applied to the original input. For an $M$-layer network acting as a classifier, an initial input vector $\bs{i}^{(1)}$ undergoes a series of transformations such that the final layer encodes an approximation to the desired output of the network,
\[
\bs{i}^{(1)} \rightarrow \bs{i}^{(2)} \rightarrow \ldots \rightarrow \bs{i}^{(M)} \mapsto \mathcal{C}(\bs{i}^{(M)}) \approx c.
\]
for some decoding function, $\mathcal{C}(\cdot)$. The $j^{\text{th}}$ element of the input vector transformed up to the $k^{\text{th}}$ layer is given by
\begin{equation}
\bs{i}_{j}^{(k)} = \phi \left( \sum_{l} W_{jl} i_l^{(k-1)} + b_j \right)
\end{equation}
where $\phi$ is a suitable non-linear function and $W_{jl}$ and $b_j$ are real, network parameters. Throughout this work we utilise the Rectified Linear Unit (ReLU) as our simple non-linear activation function, where
$
\phi_{\text{RL}} (x) = \text{max}( 0, x).
$
Such neural networks are typically optimised via a gradient descent learning scheme of some cost function $L(\bs{\theta})$ by means of backpropagation \cite{nielsenNN}, a fundamental learning mechanism that formulates the underlying efficiency and performance of neural networks. For classification problems, the cost function may take numerous forms, such as the mean classification error, or cross entropy loss etc. that we wish to minimise.
\par
Convolutional neural networks (CNNs) are a branch of network architectures that make use of convolutional layers of neurons, in which the hidden units have \textit{local receptive fields} which are tied across the input \cite{probML}. The goal is to not only reduce the number of network parameters, but more importantly to extract critical features of input space that can be reused elsewhere, allowing the network to learn key characteristics with translational invariance.
Figure \ref{fig:Classifiers}(a) depicts the generic structure of a CNN. The transformation according to a convolutional layer is described by the operation of parallel filters which can simultaneously learn unique features.\par 
Convolutional layers are ordered in a hierarchical manner, such that subsequent layers learn increasingly complex properties. For this reason, CNNs are particularly prevalent in image processing; early layers are used to identify basic, broad characteristics of images, such as edges and frames, whilst subsequent layers are then able to learn more complex structures, such as faces, hair, and texture.
Furthermore, CNNs are very effective at noise reduction, i.e.~performing effective image classification even amongst noisy inputs. In fact, noise can be shown to improve CNN classifier resilience \cite{noisyCNN}, allowing the network to better generalise to a wide array of input possibilities rather than overfit on precise image patterns. This makes them an ideal image processing tool that could be enhanced by means of quantum sensors for thermal imaging tasks.

\subsection{Quantum Enhancement}
We are now interested in determining how discrimination errors according to classical/quantum sensors propagate through to classification error of channel patterns when the statistical learning techniques of the previous section are employed.\par
Here we consider a practical classification example using the MNIST data-set \cite{mnist}, a commonly used data-set of $28\!\times\!28$ pixel images of handwritten digits 0-9, used for the benchmarking of many machine learning methods and models. The typical data-set is grey-scale, so we polarise the images such that each pixel only embodies one of two environments ${\nu}_j = \{\nu_B,\nu_T\}$. Since the channel distribution $\pi_{\bs{i}}$ is unknown and too difficult to characterise, it proposes an ideal application of data-driven techniques. Whilst this represents a toy example, it still captures the intent of thermal imaging and pattern classification. The data-set consists of a maximum training set $\mathcal{T}_{\text{max}} = \{\bs{i}_k ; c_k\}_{k}$, and an evaluation set $\mathcal{V} = \{\bs{i}_l ; c_l\}_{l}$ where $|\mathcal{T}_{\text{max}}| = 6\times10^4$ and $|\mathcal{V}| = 1\times10^4$. Using these data-sets we may quantify realistic error-rates on classification using quantum vs.~classical sensors on thermal images. \par
Bounding the single pixel error probabilities according to \cite{PirBD}
\begin{equation}
\frac{1 - \sqrt{1-F^{2M}}}{2} \leq p_\text{err}^{\text{pixel}} \leq \frac{F^M}{2},
\end{equation}
where $F$ is the fidelity between background/channel outputs as before, we may then study how these pixel errors propagate to global image classification in both quantum/classical cases. The error $E^T$ is then approximated as the expected probability of misclassification (using the NN or CNN classifier) across the evaluation set $\mathcal{V}$, using a random training subset $\mathcal{T} \subseteq \mathcal{T}_{\text{max}}$ where $|\mathcal{T}| = T$,
\begin{equation}
E^T \approx  \frac{1}{|\mathcal{V}|} \sum_{\bs{i} \in \mathcal{V}} \delta\big(\tilde{c}({\bs{i}},\mathcal{T}), c({\bs{i}})\big).
\end{equation}
Here, $\tilde{c}({\bs{i}},\mathcal{T})$ is the inferred label of $\bs{i}$ from the $T$-size training set NN/CNN classifier, $c({\bs{i}})$ is the true label, and $\delta$ is a Kronecker delta function such that 
\begin{equation}
\delta\big(\tilde{c}({\bs{i}},\mathcal{T}), c({\bs{i}})\big) = \begin{cases} 1, & \text{ if } \tilde{c}({\bs{i}},\mathcal{T}) = c({\bs{i}}), \\
0, & \text{ otherwise.} 
\end{cases}
\end{equation} 
Performing this computation using sensors with different properties, we can approximate regions within which the optimal error rate exists for finite training sets.\par
\begin{figure}[t!]
\hspace{0.4cm} (a) Classical \hspace{1.9cm} (b) Quantum  \\
\includegraphics[width=0.539\linewidth]{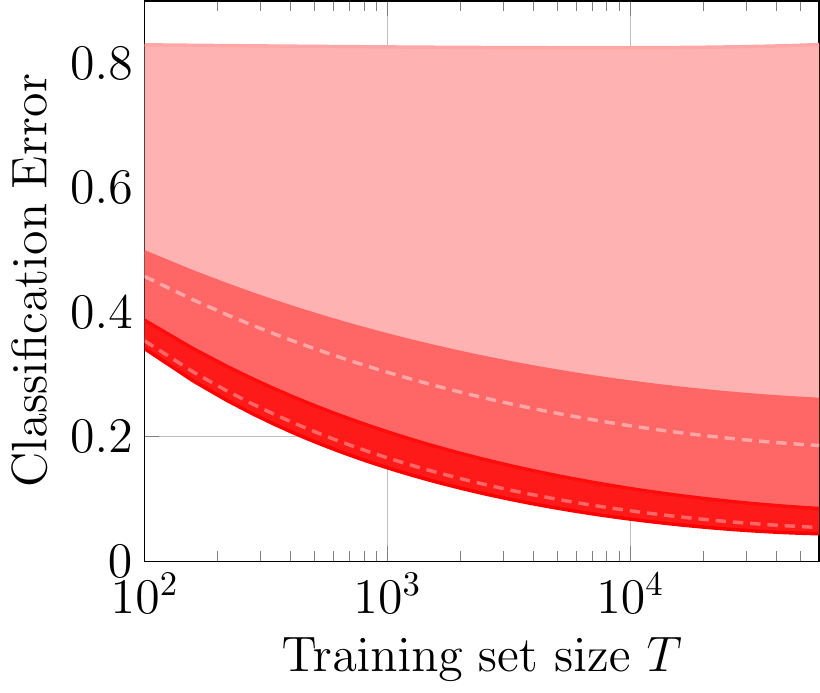}\hspace{-0.7cm}
\includegraphics[width=0.525\linewidth]{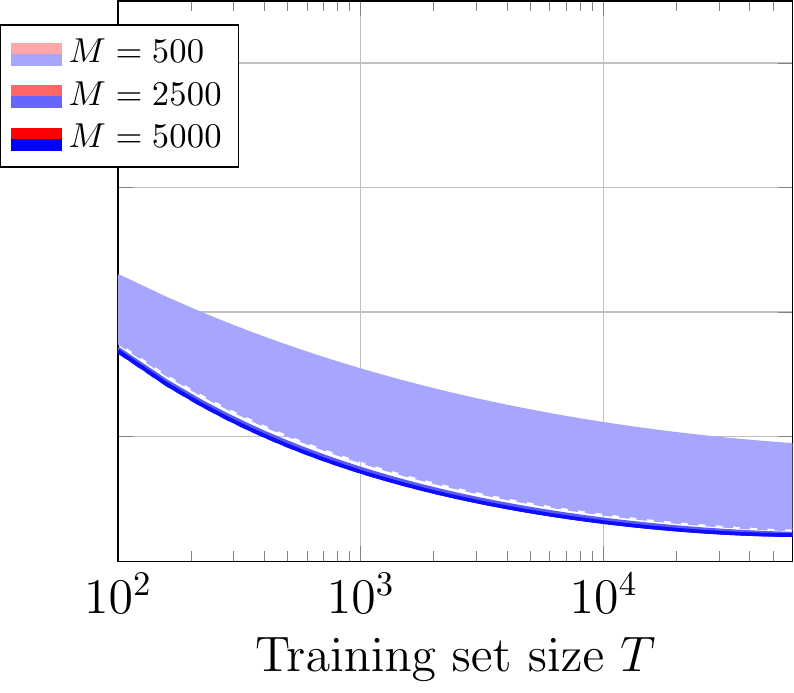}
\caption{Thermal image classification performance on a thermally rendered MNIST handwritten digit data-set, using a NN Classifier. Each pixel is modelled as a thermal-loss channel with $\{\epsilon_B,\epsilon_T\} = \{18.5, 20.2\}$. 
The error region for the quantum-enhanced classifier at $M=5000$ is extremely narrow, and essentially takes the form of the lower bound for $M=2500$. Dashed lines indicate lower bound of the correspondingly coloured error region.} 
\label{fig:NNclassifier}
\end{figure}

 \begin{figure}[t!]
\hspace{1.4cm}(a)\\
\includegraphics[width=0.8\linewidth]{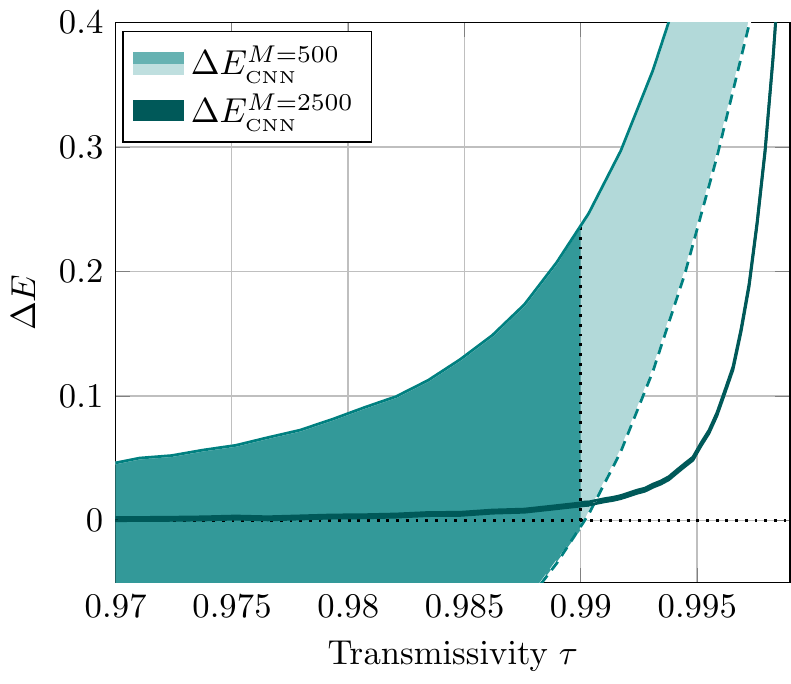}\\
\hspace{1.4cm}(b)\\
\includegraphics[width=0.85\linewidth]{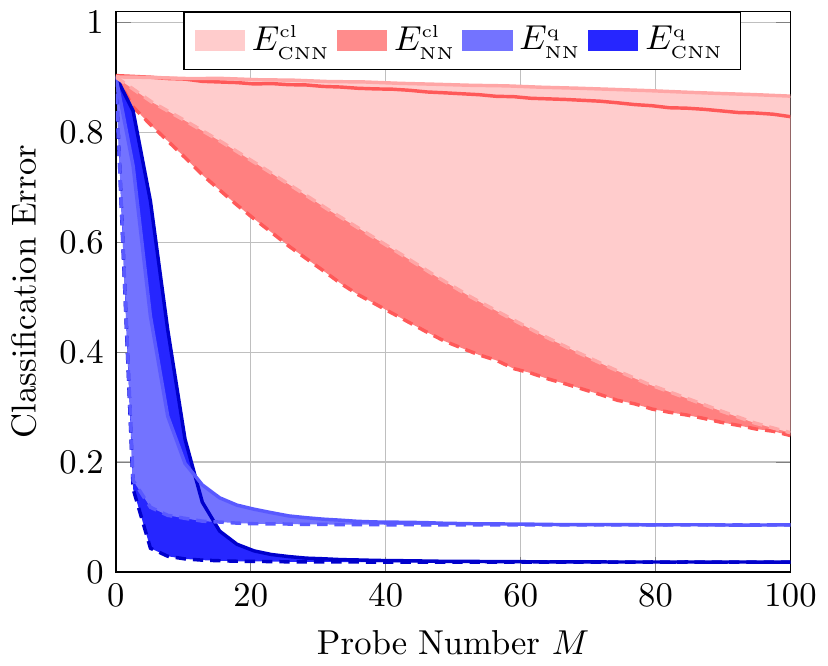}\hspace{-0.3cm}
\caption{Quantum-enhanced thermal imaging on the MNIST data set, using NN and CNN pattern classifiers, for (a) thermal-loss channel patterns with $\{\epsilon_B,\epsilon_T\} = \{18.5, 20.2\}$ and (b) Additive noise channels with $\{\nu_B,\nu_T\} = \{0.02,0.01\}$. Panel (a) depicts quantum advantage bounds for thermal-loss channels for variable $\tau$, using a CNN classifier. Change in colour of the regions for $M=500$ is used to indicate a transition from potential advantage to guaranteed advantage at $\tau \sim 0.99$ using realistic quantum resources. Panel (b) illustrates the error-classification bounds for additive noise channels under variable probe copy number $M$, using both statistical classifier models. In both cases $T=10000$, dashed lines indicate lower bounds, and performances have been averaged over a number of experiments.}
\label{fig:CNNvsNN}
\end{figure}
Figures \ref{fig:NNclassifier} and \ref{fig:CNNvsNN} describe the results of our numerical simulations, analysing the error rates for both the NN and CNN classifiers for the thermal-loss and additive noise channels. In Fig.~\ref{fig:NNclassifier} (a) and (b) the performance of classical and quantum sensors respectively are investigated for the NN classifier and thermal-loss channels; plotting regions within which the error rate may lie (for numerous probe copy numbers $M$) against the dimension of the training set. 
These error ranges were numerically deduced through an empirical finite-sample risk analysis on an evaluation set using the NN classifier, for different sizes of training set $T$ and $\tau = 0.99$, carried out for an optimal classical sensor (a) and a quantum sensor (b). The behaviour with respect to training set dimension was then interpolated via Eq.~(\ref{eq:Snapp}) (truncated at $j=5$). Importantly, in the regime of low-loss quantum sources offer a more confident error rate, such that the separation between the upper and lower bounds is much narrower. As one would expect, increasing the training set dimension leads to significantly lower error rates, as the classifier becomes increasingly more likely to find a neighbouring sample within its training set that will provide a correct classification. \par
However, we see from Fig.~\ref{fig:CNNvsNN} (a), quantum advantage for thermal-loss channels is very fragile with respect to loss. Here we focus on the CNN classifier, which is superior in its classification power (however this behaviour is synonymous with the NN classifier). We plot the regions $$
\left[ \Delta E^\text{min},  \Delta E^{\text{max}} \right]  \defeq \left[ E_{\text{cl}}^{\text{L}} -  E_{\text{q}}^{\text{U}}, E_{\text{cl}}^{\text{L}} -  E_{\text{q}}^{\text{L}} \right]
$$ (for the respective probe numbers) which are analogous definitions of MGA and MPA from Eqs.~(\ref{eq:MGA})-(\ref{eq:MPA}). We observe that even for transmissivities as high as $\tau \sim 0.97$ and a sufficient number of probe-copies, the advantage is either negligible or insufficiently guaranteed; only as $\tau$ approaches unity does advantage become guaranteed in relevant magnitudes (irrespective of the classifier). These results motivate the use of quantum sources only in a very low-loss regime (concluded previously), as the decoherence involved with such channels removes advantageous entanglement between idler and signal probe too quickly. \par
As has been observed in previous sections, we can then expect much greater advantage associated with additive noise channels. Indeed, Fig.~\ref{fig:CNNvsNN}(b) shows the average classification error for additive noise channels for both classifiers with comparable resources $T = 10000$. Now we observe a significant level of quantum advantage, for a very low number of probes.\par
It is interesting to note that while both the CNN and NN classifiers follow very similar behaviours with respect to channel properties, there are a few key differences. In particular, provided with sufficient resources the CNN will converge to a superior misclassification rate than the NN classifier, clearly visible in Fig.~\ref{fig:CNNvsNN}(b). The simplicity of the NN approach means that its performance is wholly dependent on the dimension of its training set, and the precision of the sensor. Meanwhile, the CNN is capable of learning more intricate and detailed image features, meaning that it can more efficiently exploit the information provided from its training set. \par
Conversely, when the pixel error rate is very low, the CNN will typically perform slightly worse. This can be seen in the error regions of the classical sensors in Fig.~\ref{fig:CNNvsNN}(b). If the pixel error rate is too high, the CNN will struggle to extract any consistent features from a sampled pattern, making classification very difficult. Therefore in a high noise regime, NN classifiers may display improved robustness.

\section{Discussion}
Following from \cite{PatternRecog, OptEnvLoc, ULMCD} we have formulated a problem setting for thermal imaging wherein the classification of an image is equated to the multi-channel discrimination of bosonic Gaussian channels. By considering binary collections of equally transmissive thermal-loss channels and additive noise channels, we exploited their joint teleportation covariance and applied the technique of teleportation stretching to derive ultimate lower bounds on the error probability of classification. This allowed for the numerical and analytical analysis of these bounds for a number of image spaces, (uniform and non-uniform) under the formulation of channel position finding. These results emphasised the underlying quantum advantage that can be achieved in the limit of low-loss thermal channel patterns.  Furthermore, we considered quantum-enhanced statistical pattern recognition for thermal imaging, making use of relevant machine learning methods in order to investigate the potential enhancement offered by quantum sources in a broader image setting. \par
Indeed, the limitation of entanglement enhanced discrimination resides in the susceptibility of such states to decoherence, which is particularly prominent in the presence of both loss and noise. While the discrimination of pure-loss and additive-noise channels can be impressively enhanced by quantum sources, the combination of these effects can dramatically degrade any guaranteed advantage. \par
Nonetheless, low-loss is ensured when the targets are limited to a short-range, making it easier to preserve quantum correlations between signal and idler-modes. This regime is realistic and important in future sensing applications, such as in low energy, non-invasive biomedical scanning. Alternative applications exist in communications, where eavesdropper activity along low-loss, multimode communication lines can be identified via their environmental noise properties. Eavesdroppers can be discriminated using appropriate environment localisation and pattern recognition techniques, such as those explored in this paper. These fundamental limits affirm the benefits and prospects of short-range, quantum-enhanced thermal imaging.

\begin{acknowledgements}
C.H acknowledges funding from the EPSRC via a Doctoral Training Partnership (EP/R513386/1). S.P acknowledges funding from the European Union’s Horizon 2020 Research and Innovation Action under grant agreement No.~862644 (Quantum readout techniques and technologies, QUARTET). L.B.~acknowledges support by the program ``Rita Levi Montalcini" for young researchers. 
\end{acknowledgements}

\appendix

\section{Probe State Fidelities}
\subsection{Choi State Fidelities}
The covariance matrix of a finite-energy Choi state of a Gaussian phase insensitive channel is given by
\begin{equation}
V = \begin{pmatrix} a {I} & \sqrt{\tau(a^2 - 1/4)}Z \\
\sqrt{\tau(a^2 - 1/4)}Z & (a\tau + \nu) {I}
\end{pmatrix},
\end{equation}
where $
a = \bar{n}_S + 1/2
$ is the squeezing parameter of $\Phi_{a}$, defined by the average mean photon number $\bar{n}_S$ per mode. It is now possible to evaluate the fidelities between Choi states of these channels via multi-modal fidelity formulae from \cite{GFid}. Consider the fidelity between Choi-states of thermal-loss/amplifier channels with parameters $\{\tau, \epsilon_T, \epsilon_B\}$ where $\epsilon_j = \bar{n}_j + 1/2$ are the local mean photon numbers for $j = B,T$. Taken in the limit of infinite-squeezing ($a \rightarrow \infty$), the fidelity between the Choi matrices depends only on the thermal parameters of each channel and not on $\tau$ \cite{OptEnvLoc},
\begin{equation}
F_{\text{th}}(\epsilon_{T},\epsilon_{B}) = \frac{\sqrt{\epsilon_T \epsilon_B + 1 + \sqrt{(4\epsilon_T^2 -1)(4\epsilon_B^2 - 1)}}}{\sqrt{2}(\epsilon_T + \epsilon_B)}. \label{eq:ThermChoiFid}
\end{equation}
For Gaussian additive-noise channels with noise parameters $\{\nu_T, \nu_B\}$, the fidelity in the limit of infinite squeezing takes a more succinct form and we get
\begin{equation}
F_{\text{add}} ({\nu_T, \nu_B}) = \frac{2\sqrt{\nu_T \nu_B}}{\nu_T + \nu_B}. \label{eq:AddChoiFid}
\end{equation}

\subsection{Optimal Classical State Fidelities}
The optimal classical strategy for environment localisation may simply make use of a block-protocol using $M$-copy vacuum states $\ket{0}^{\otimes M}$ in order to probe a channel pattern. For thermal-loss/amplifier channels, the optimal classical strategy provides an output fidelity \cite{OptEnvLoc},
\begin{align}
&F_{\text{th}}^{\text{cl}} (\tau,\epsilon_T, \epsilon_B) = \frac{
\sqrt{\alpha + \delta} + \sqrt{\alpha - \beta}}{\beta}, \\
&\alpha = 4\epsilon_T\epsilon_B |1-\tau|^2 + 2(\epsilon_T+\epsilon_B)\tau|1-\tau| + (1+\tau^2),\nonumber \\
&\beta = 2(\tau + (\epsilon_T +\epsilon_B)|1-\tau|).\nonumber
\end{align}
For additive noise channels the optimal classical output fidelity is given by,
\begin{equation}
F_{\text{add}}^{\text{cl}} = \frac{1}{\sqrt{(\nu_T +1)(\nu_B +1)} - \sqrt{\nu_T \nu_B}}
\end{equation}
These fidelities can then be used within error bounds to characterise the optimal classical performance.

\begin{figure}[t!]
\includegraphics[width=\linewidth]{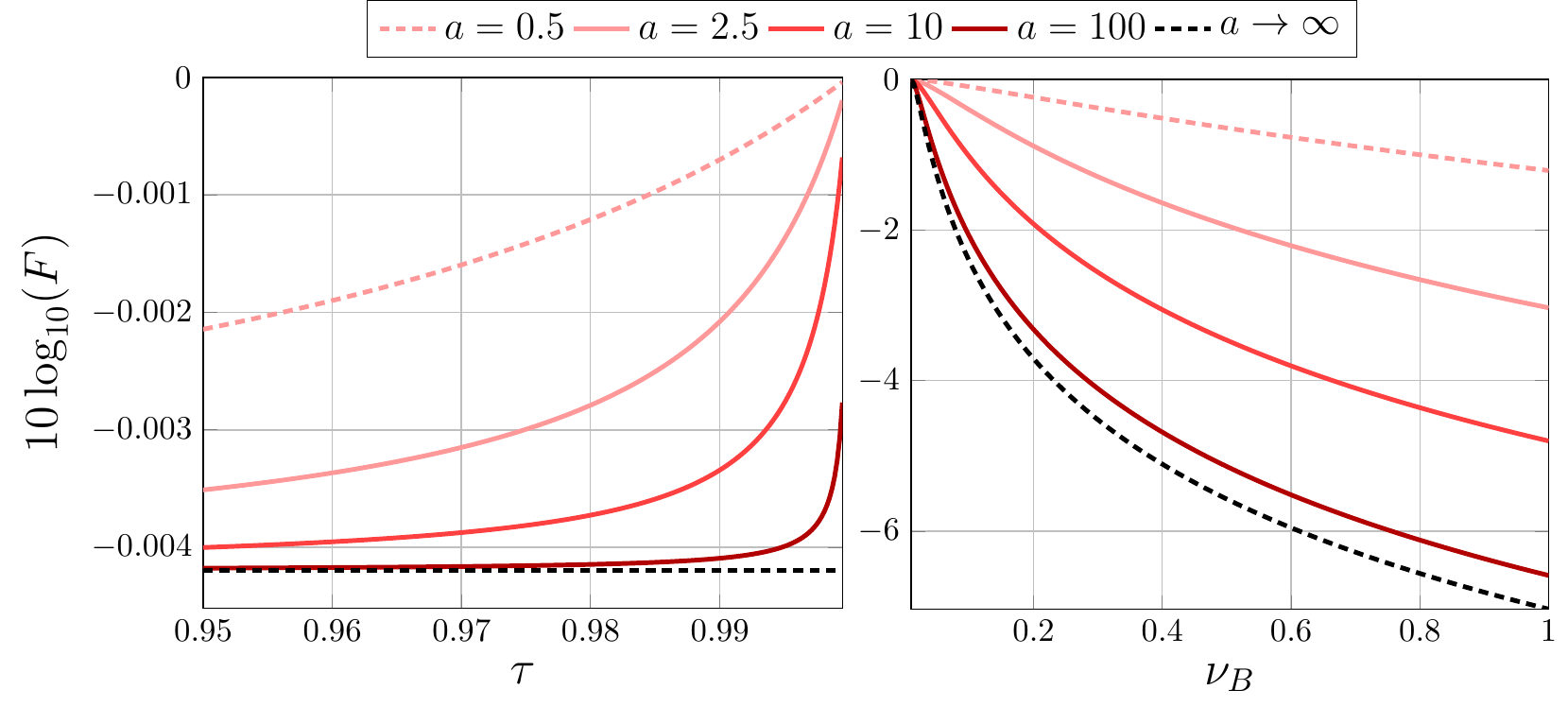}\\
\hspace{9mm} (a) Thermal-Loss \hspace{1.2cm} (b) Additive-Noise
\caption{Finite-energy output fidelity behaviour for (a) thermal-loss channels with background/target parameters $\{\epsilon_B,\epsilon_T\} = \{18.5,20.2\}$, and (b) additive-noise channels with target noise $\nu_T = 0.01$. Beginning in a vacuum state (optimal classical probe) the finite-energy Choi-state approximations quickly approach the optimal quantum probe state fidelities in the limit of infinite squeezing, for realistic energetic resources, shown here for $a \in \{2.5,10,100\}$. }
\label{fig:FEProbes}
\end{figure}

\section{Finite-Energy Considerations\label{sec:FEFids}}
In this work we showed that TMSV states in the limit of infinite-squeezing derive ultimate limits for multi-channel discrimination of telecovariant quantum channels. However, it is not immediately clear how realistic probe states (which necessarily undergo finite-squeezing) approach this ultimate limit, and what the energy requirements are to observe meaningful advantage. We can readily access this information by looking at the finite-energy formats of the Choi-state fidelities in Eqs.~(\ref{eq:ThermChoiFid})-(\ref{eq:AddChoiFid}). By studying the behaviour of these fidelities with respect to increasing energy, one can observe the resources required to realistically approach the ultimate limits derived in the main text.\par
Contrary to Eq.~(\ref{eq:ThermChoiFid}), the finite-energy Choi-state fidelity associated with thermal-loss channels does depend on the transmissivity, given by
\begin{align}
&F_{\text{th}}^{a} (\tau,\epsilon_T, \epsilon_B) = \frac{\sqrt{2}\left(
\sqrt{\xi + \omega} + \sqrt{\xi- \omega}\right)}{\omega},\\
&\xi = 4\epsilon_T\epsilon_B + 4a^2(4\epsilon_T\epsilon_B +1) + |1-\tau|^2(4a^2-1) \\
&\times \sqrt{(4\epsilon_T^2 -1)(4\epsilon_B^2 - 1)}  + 8a\tau|1-\tau|(1+\tau)^2(\epsilon_T+\epsilon_B), \nonumber \\
&\omega = 4(\tau + (\epsilon_T +2a\epsilon_B)|1-\tau|).
\end{align}
Meanwhile, the finite-energy Choi-state fidelity associated with probing additive-noise channels admits the following form,
\begin{align}
&F_{\text{add}}^{a} ({\nu_T, \nu_B}) = \frac{2a\sqrt{\nu_T\nu_B} + \sqrt{(2a\nu_T+1)(2a\nu_B+1)}}{2a(\nu_T+\nu_B)+1}.
\end{align}
In both cases the optimal classical and quantum fidelities can be recovered by taking the correct energy limits respectively,
\begin{gather}
F_j  = \lim_{a \rightarrow \infty} F_j^a,\hspace{4mm} F_j^{\text{cl}}  = \lim_{a \rightarrow \frac{1}{2}} F_j^a,
\end{gather}
where the subscript $j \in \{\text{th},\text{add}\}$ is used to label either thermal-loss or additive-noise fidelities. Furthermore, recall that $a \defeq \bar{n}_S + \frac{1}{2}$, where $\bar{n}_S$ is the mean photon energy of incident probe states.
\par

Figure~\ref{fig:FEProbes} illustrates how the finite-energy fidelities approach the ultimate limits. For thermal-loss channels of high-transmissivity in Fig.~\ref{fig:FEProbes}(a), we show that the fidelity gap is quickly closed by finitely-squeezed probe states, and that even a small amount of squeezing $a = 2.5$ ($\bar{n}_S = 2$) leads to a significant improvement over the classical output fidelity. For larger, but very much realistic squeezing $a = 100$, the ultimate limit is essentially saturated for $\tau \lesssim 0.99$.\par
A similar behaviour can be observed for additive-noise channels in Fig.~\ref{fig:FEProbes}(b), where the gap between the optimal classical and quantum output fidelities is much more pronounced. Nonetheless, this fidelity separation is rapidly traversed for realistic energetic resources, emphasising the feasibility of high performance, entangled probe state protocols.

\section{Temperature of Thermal Environments\label{sec:Temps}}
Throughout this paper we characterise thermal environments via their mean photon number using $\bar{n}_j$/$\epsilon_j$. Using a Boltzmann distribution these can be converted to temperatures that characterise background and target pixels of thermal images. Assuming that channels are probed using microwave radiation of wavelength $\lambda \sim 1\text{mm}$, then we find pixel temperatures (in Kelvin),
\begin{equation}
T_j=  \frac{hc}{k\lambda  \log\left(\frac{1}{\bar{n}_j} + 1\right)}.
\end{equation}
where $k$ is Boltzmann's constant, $h$ is Planck's constant, and $c$ is the speed of light.

\section{Uniform Image Spaces\label{sec:AppUni}}
Probing each pixel with $M$ copies of a chosen input state provides the ultimate lower bound (for uniform \textit{a priori} probabilities) given by,
\begin{align}
\frac{1}{2^{2m +1}} \sum_{\bs{i} \neq \bs{i}^{\prime}} F^{2M}(\rho_{\bs{i}}, \rho_{{\bs{i}^\prime}} ) \leq p_{\text{err}} \leq \frac{1}{2^{m}} \sum_{\bs{i} \neq {\bs{i}^\prime}} F^{M}(\rho_{\bs{i}}, \rho_{{\bs{i}^\prime}} ).
\end{align}
Since we are dealing with telecovariant channels, we fundamentally bound these error probabilities through teleportation stretching. Stretching the discrimination protocol by supplanting the respective  output states with the Choi states of each channel,
$
{\rho}_{\mathcal{E}_{\bs{i}}^m} = \bigotimes_{k=1}^m \rho_{\mathcal{E}_{\bs{i}_k}}
$
such that $i_k \in \{ B,T \}$, the fidelity between any two distinct, unequal patterns will be given by
\begin{equation}
F({\rho}_{\mathcal{E}_{\bs{i}}}, {\rho}_{\mathcal{E}_{{\bs{i}^\prime}}}) = F({\rho}_{\mathcal{E}_{B}}, {\rho}_{\mathcal{E}_{T}} )^{\text{hamming}(\bs{i},{\bs{i}^\prime})}.
\end{equation}
Considering the quantity,
\begin{equation}
D_m(f) = \frac{1}{2^m} \sum_{\bs{i}\neq {\bs{i}^\prime}} f^{\text{hamming}(\bs{i},{\bs{i}^\prime})} =  (f+1)^m - 1,  \label{eq:BcodeHam}
\end{equation}
we find a closed form for this quantity. This can be proven via recursion as in \cite{PatternRecog}, or by rewriting $\sum_{\bs{i}\neq \bs{i}^{\prime}} = \sum_{\bs{i}, \bs{i}^{\prime}} - \sum_{\bs{i}=\bs{i}^{\prime}}$ and using multiplicativity of the fidelity. Using $D_m(f)$ the ultimate lower bound in Eq.~(\ref{eq:LB_Bcode}) can then be shown. An upper bound using Eq.~(\ref{eq:JointUB}) can also be derived in this manner, such that 
\begin{equation}
p_\text{err} \leq  [ F^{M}(\rho_{\mathcal{E}_{B}}, \rho_{\mathcal{E}_{T}}) + 1]^m - 1, \label{eq:UB_Bcode}
\end{equation} 
however this was shown to always be looser than Eq.~(\ref{eq:LocalM}).

\section{Channel Position Finding\label{sec:CPF}}
\subsection{$k$-CPF Bounds}
Let ${\mathcal{U}_{{\textsf{\tiny CPF}}}^k}$ be the set of $m$ pixel channel patterns with strictly $k$-targets, such that $|{\mathcal{U}_{{\textsf{\tiny CPF}}}^k}| = {m\choose k}$ (where ${m\choose k}$ is the binomial coefficient).  We wish to compute the sum
\begin{equation}
D_m^k = \frac{1}{{m\choose k}} {\sum_{ (\bs{i}\neq\bs{i}^\prime) \in{\mathcal{U}_{{\textsf{\tiny CPF}}}^k} }} f^{\text{hamming}(\bs{i},\bs{i}^\prime)}.
\end{equation}
For $\bs{i},\bs{i}^\prime \in {\mathcal{U}_{{\textsf{\tiny CPF}}}^k}$ all possible Hamming distances take the values:
\begin{equation}
\text{hamming}(\bs{i}, \bs{i}^\prime) = 2(t - k), \> \forall t \in \{k+1,\ldots, 2k\}.
\end{equation}
for maximum distance $k$ (where all $k$ target pixels are different) and minimum distance $2$ (where only one pixel is different). When $k=1$ we reduce to CPF as usual. \par
Consider each $\bs{i} \neq \bs{i}^\prime$; there exist $t \in \{k+1,\ldots, 2k\}$ locations in total for us to insert our target channels, and $m\choose t $ ways for this to occur. We may then place $k$ targets into one of the channel patterns $\bs{i}$, with $t\choose k$ ways to do this. Having used $k$ of our $t$ locations, there now only exist $k-(t-k)$ channels that can be freely placed into the second pattern $\bs{i}^\prime$, with $k\choose 2k-t$ ways to do this. Via this counting argument the above sum can be rephrased,
\begin{align}
{\sum_{ (\bs{i}\neq\bs{i}^\prime) \in{\mathcal{U}_{{\textsf{\tiny CPF}}}^k} }} &f^{\text{hamming}(\bs{i},\bs{i}^\prime)} = \sum_{t=k+1}^{2k} {m\choose t}{t\choose k} {k \choose 2k-t} f^{2(t-k)}, \nonumber \\
&= {m\choose k} \left[ {{}_2 F_1}(-k,k-m,1,f^2) - 1 \right],
\end{align}
where ${}_2F_1$ is the standard hypergeometric function. Now we can define optimal quantum upper bounds and ultimate lower bounds on thermal $k$-CPF pattern recognition. Given
\begin{equation}
D_m^k (f) = \left[ {{}_2 F_1}(-k,k-m,1,f^2) - 1 \right] \label{eq:kCPFanalytic},
\end{equation}
then the bounds, assuming uniform \textit{a priori} probabilities $\pi_{\bs{i}} = {m\choose k}^{-1}$, can be written as,
\begin{align}
p_{\text{err}}^{\text{opt}} &\geq \frac{D_m^k \left[ F^{2M}(\rho_{\mathcal{E}_T},\rho_{\mathcal{E}_B}) \right]}{2{m\choose k}},\\
p_{\text{err}}^{\text{opt}} &\leq D_m^k \left[ F^{M}(\rho_{\mathcal{E}_T},\rho_{\mathcal{E}_B}) \right].
\end{align}

\subsection{$k$-Bounded CPF Bounds \label{sec:kBCPF}}
If the prior distribution of target channels is not uniform, then we may consider $k$-BCPF, such that there exists an image space $\mathcal{U}_{\textsf{\tiny CPF}}^{\bs{k}} = \bigcup_{k\in\bs{k}} \mathcal{U}_{\textsf{\tiny CPF}}^k$ such that $\bs{k}$ contains all possible numbers of target channels in any image in the space. In order to derive error probability bounds we must solve,
\begin{equation}
D_m^{\bs{k}}(f) = \sum_{ (\bs{i}\neq\bs{i}^\prime) \in{\mathcal{U}_{\textsf{\tiny CPF}}^{\bs{k}}} } f^{\text{hamming}(\bs{i},\bs{i}^\prime)} . \label{eq:Boundk}
\end{equation}
The property $\bs{i}\neq\bs{i}^\prime$ only applies when each image is being drawn from the same set ${\mathcal{U}_{{\textsf{\tiny CPF}}}^k}$, otherwise they are unequal by construction, hence we 
define the following quantity for $k\neq l$
\begin{equation}
\tilde{D}_m^{k,l}(f) =  {\sum_{\bs{i} \in {\mathcal{U}_{{\textsf{\tiny CPF}}}^k}, \bs{i}^\prime \in {\mathcal{U}_{{\textsf{\tiny CPF}}}^l} }} f^{\text{hamming}(\bs{i},\bs{i}^\prime)},
\end{equation}
which allows us to rewrite Eq.~(\ref{eq:Boundk}) by splitting it into independent diagonal and off diagonal summations,
\begin{equation}
D_m^{\bs{k}} (f) = \sum_{j \in \bs{k} } {D_{m}^j(f)}+  \sum_{ i \in \bs{k} } \sum_{l \neq i} {\tilde{D}_{m}^{i,l}(f)} \label{eq:Boundk3}.
\end{equation}
These $\tilde{D}_m^{k,l}(f)$ represent off diagonal functionals, in the sense that they count contributions from dissimilar $k$-CPF sets. Given $\bs{i} \in {\mathcal{U}_{{\textsf{\tiny CPF}}}^k}$ and $\bs{i}^\prime \in {\mathcal{U}_{{\textsf{\tiny CPF}}}^l}$ and assuming that $k < l$, all possible Hamming distances take the values:
\begin{equation}
\text{hamming}(\bs{i}, \bs{i}^\prime) = 2t - (k+l), \> \forall t \in \{l,\ldots, k+l\},
\end{equation}
where the minimal distance is $l-k$ (all of $\bs{i}^\prime$'s targets align with $\bs{i}$'s) and maximal distance is $k+l$ (the least possible amount of alignment between the patterns). Applying an identical counting argument as for standard $k$-CPF, one can express the off diagonal sums as
\begin{equation}
\tilde{D}_m^{k,l}(f) = \sum_{t=l}^{k+l} {m\choose t} {t\choose l} {l \choose (k+l)-t} f^{2t - (k+l)}.
\end{equation}
In order to retrieve the original case of $k=l$, we simply adjust the $t$ domain to eliminate self similar terms i.e.\ the minimum value of $t$ becomes $t=k+1$. This is once again solved by the standard hypergeometric function as
\begin{equation}
\tilde{D}_m^{k,l}(f) = {m\choose l} {l\choose k} f^{k-l} {}_2 F_1 (-k, l-m, l-k+1, f^2).
\end{equation}
Using this result one can write optimal upper and lower bounds for $k$-BCPF using Eq's. \ref{eq:Boundk} and \ref{eq:Boundk3}, in conjunction with uniform weights,
$
\pi_{\bs{k}} = \left[ \sum_{i \in \bs{k} }  {m\choose i} \right]^{-1},
$
then
\begin{align}
p_{\text{err}}^{\text{opt}} &\geq \frac{\pi_{\bs{k}}^2}{2}  D_m^{\bs{k}} \left[ F^{2M}(\rho_{\mathcal{E}_T},\rho_{\mathcal{E}_B}) \right], \\
p_{\text{err}}^{\text{opt}} &\leq {\pi_{\bs{k}}}  D_m^{\bs{k}} \left[ F^{M}(\rho_{\mathcal{E}_T},\rho_{\mathcal{E}_B}) \right].
\end{align}
These bounds represent ($i$) the ultimate lower bounds on thermal pattern classification \textit{given} that the number of target channels is contained in $\bs{k}$, and ($ii$) a quantum upper bound on the optimal error probability based on this pattern assumption and joint measurements. When $|\bs{k}|=m$ (contains all possible target numbers) we reproduce bounds for uniform channel patterns, and when $|\bs{k}|=1$ (contains only one target number) we reduce to $k$-CPF.\par

\end{document}